\documentclass[aps,10pt,a4paper,showpacs,pre,twocolumn,floatfix]{revtex4-1}
\usepackage{amssymb}
\usepackage{mathrsfs}
\usepackage{stmaryrd}
\usepackage{subfigure}

\usepackage{graphicx}
\usepackage{color}

\usepackage{fancyhdr}
\newcommand{\mb}{\mathbf}
\newcommand{\mc}{\mathcal}

\newcommand{\mr}{\mathrm}
\newcommand{\tb}{\textbf}

\newcommand{\etal}{et al.\ }

\newcommand{\ud}{{\mr d}}
\newcommand{\rr}{{\mb r}}
\newcommand{\qq}{{\mb q}}
\newcommand{\LL}{\Lambda}
\newcommand{\LLD}{\Lambda^{\mbox{\tiny local}}}
\newcommand{\LCD}{\Lambda^{\mbox{\tiny chain}}}

\newcommand{\tLL}{\tilde{\Lambda}}
\newcommand{\tLLD}{\tilde{\Lambda}^{\mbox{\tiny local}}}
\newcommand{\tLCD}{\tilde{\Lambda}^{\mbox{\tiny chain}}}
\newcommand{\tLDB}{\tilde{\Lambda}^{\mbox{\tiny Debye}}}
\newcommand{\tLNL}{\tilde{\Lambda}^{\mbox{\tiny nonlocal}}}

\newcommand{\ab}{{\alpha \beta}}
\newcommand{\ga}{{\alpha}}
\newcommand{\gb}{{\beta}}
\newcommand{\ff}{{\tilde{\mathbf{F}}}}

\begin{document}

\title{Dynamic density functional theories for inhomogeneous polymer systems compared to Brownian dynamics simulations}

\author{Shuanhu Qi}
\affiliation{Institut f\"{u}r Physik, Johannes Gutenberg-Universit\"{a}t Mainz, Staudingerweg 7, D-55099 Mainz, Germany}

\author{Friederike Schmid}
\affiliation{Institut f\"{u}r Physik, Johannes Gutenberg-Universit\"{a}t Mainz, Staudingerweg 7, D-55099 Mainz, Germany}

\bigskip

\begin{abstract}

Dynamic density functionals (DDFs) are popular tools for studying the dynamical
evolution of inhomogeneous polymer systems. Here, we present a systematic
evaluation of a set of diffusive DDF theories by comparing their predictions
with data from particle-based Brownian dynamics (BD) simulations for two
selected problems: Interface broadening in compressible A/B homopolymer
blends after a sudden change of the incompatibility parameter, and microphase
separation in compressible A:B diblock copolymer melts.  Specifically, we
examine (i) a local dynamics model, where monomers are taken to move
independently from each other, (ii) a nonlocal ``chain dynamics" model, where
monomers move jointly with correlation matrix given by the local chain
correlator, and (iii,iv) two popular approximations to (ii), namely (iii) the
Debye dynamics model, where the chain correlator is approximated by its value
in a homogeneous system, and (iv) the computationally efficient ``external
potential dynamics" (EPD) model. With the exception of EPD, the value of
the compressibility parameter has little influence on the results. In the
interface broadening problem, the chain dynamics model reproduces the BD data
best. However, the closely related EPD model produces large spurious artefacts.
These artefacts disappear when the blend system becomes incompressible. In
the microphase separation problem, the predictions of the nonlocal models
(ii-iv) agree with each other and significantly overestimate the ordering time,
whereas the local model (i) underestimates it.  We attribute this to the
multiscale character of the ordering process, which involves both local and
global chain rearrangements.  To account for this, we propose a mixed
local/nonlocal DDF scheme which quantitatively reproduces all BD simulation
data considered here.

\end{abstract}

\maketitle

\newpage

\section{Introduction}

The dynamics of phase transitions and morphology formation in inhomogeneous
polymeric systems is of fundamental as well as great practical interest
\cite{multiphase_book, morphology_book,scattering_spinodal_decomposition,
neutron_interface_formation, optoelectronics_review, micelle_fusion}. Due to
the large length and time scales on which these processes take place,
simulation studies usually rely on coarse-grained models. Among these, polymer
density functional based models have proven to be particularly useful tools for
studying structure formation on mesoscopic scales \cite{Kawasaki1, Kawasaki2,
Harden, dynamic_MF, EPD1997, Doi_adsorption, Kawakatsu, Shi_interface, Shima,
fluctuation_dynamics, EPD_vesicle, DFT_vesicle, spinodal_critical}.  Compared
to phase field models \cite{book_cmp, dynamic_review, ohta_kawasaki, gomez_15,
abate_16, simon_16}, they have the advantage that they retain some information
on the molecular architecture, while still treating the system at the level of
a continuum theory. 

The static polymer density functional theory is equivalent to the
``self-consistent field theory" for polymers, which can formally be ``derived"
from a particle-based model \cite{helfand_75, freed_95, review_scf} and is one
of the most successful theories of inhomogeneous polymer systems at equilibrium
\cite{matsen_review,review2_scf}. Unfortunately, a similarly systematic
construction of dynamic density functional theories is far from trivial. For
fluids made of simpler units, dynamical evolution equations for the densities
have been derived \cite{Dean1996, SDFT, DFT_fluid, DFT_spinodal, DFT_freezing,
DDFT_SD, power_functional, power_functional2} and validated by comparison with
molecular dynamics simulations \cite{DFT_fluid, DFT_JPCM}. In polymer systems,
the situation is complicated by the chain connectivity. Systematic attempts to
derive dynamic mean field theories at least for the Rouse regime
\cite{dynamics_PI_SCMF, dynamics_path_integral} have typically resulted in
approaches that are very similar to the popular ``single chain in mean field"
schemes (SCMF) \cite{ganesan1, interface_SCFBD,scmf1,scmf2,hybrid_MD}, where
one considers the time evolution of the probability distribution of whole
chains in (time-dependent) external fields. Using such schemes to further
derive explicit equations for the time evolution of the density is a formidable
challenge. Due to the wide range of time scales involved in chain dynamics, one
would expect such equations to contain memory kernels \cite{Zwanzig,
projection1, projection2}, which are difficult to handle in practice.

A popular pragmatic alternative is to construct a heuristic dynamic
density functional (DDF) Ansatz of the form \cite{Kawasaki1, Kawasaki2, Harden,
dynamic_MF, EPD1997, Doi_adsorption, Kawakatsu, Shi_interface,  Shima,
fluctuation_dynamics}
\begin{equation}
\label{eq:onsager_general}
\partial_t \rho_\ga (\rr,t)= \nabla \int \ud \rr' \LL_\ab (\rr,\rr',t) \nabla' \mu_\gb(\rr',t).
\end{equation}
Here $\rho_\ga$ is the number density field of the $\ga$th component, ($-\nabla
\mu_\gb$) is the thermodynamic driving force acting on component $\gb$ which is
derived from a Helmholtz free energy functional ${\mc F}[\{\rho\}]$ via
$\mu_\ga(\rr) = \delta {\mc F}/\delta \rho_\ga(\rr)$, and $\LL_\ab(\rr,\rr')$
is a mobility matrix which relates the density current of monomers $\ga$ at
position $\rr$ to the thermodynamic force acting on monomers $\beta$ at
position $\rr'$. We note that Eq.\ (\ref{eq:onsager_general}) describes a
diffusive (overdamped) evolution of locally conserved density fields
$\rho_\ga$, therefore it belongs to the class of model B systems
\cite{dynamic_review,book_cmp}. 

In the framework of the dynamical models of type Eq.~(\ref{eq:onsager_general}), 
the simplest approach is to postulate local coupling \cite{dynamic_MF,
Doi_adsorption,Shi_interface}, i.e., to assume that monomers move independently
from each other with a diffusion constant $D_{\mr o}$: 
\begin{equation}
\LL_\ab(\rr,\rr') = D_{\mr o} \: \rho_\ga(\rr) \: 
   \delta_\ab \: \delta(\rr-\rr') =: \LLD_\ab 
\label{eq:onsager_local}
\end{equation}
We will refer to this Ansatz as ``local dynamics''. 

Alternatively, based on Rouse dynamics and a local equilibrium assumption,
Maurits \etal \cite{EPD1997} proposed a nonlocal coupling Ansatz 
\begin{equation}
\LL_\ab(\rr,\rr') = D_{\mr c} \: P_\ab(\rr,\rr') =: \LCD_\ab,
\label{eq:onsager_chain}
\end{equation}
where $P_\ab(\rr,\rr')$ is the pair density of monomers $\ga, \gb$ from the
same chain at positions $\rr$ and $\rr'$, and $D_{\mr c}$ the diffusion
constant of whole chains. Hence, chains are assumed to move as a whole. We will
refer to this scheme as ``chain dynamics''. 

In practice, the numerical integration of the DDF equation
Eq.~(\ref{eq:onsager_general}) with the exact chain correlator
Eq.~(\ref{eq:onsager_chain}) is not straightforward, and therefore, further
approximations are usually made. For example, the two body correlator is
sometimes approximated by the corresponding correlator in the homogeneous melt
\cite{EPD1997}, i.e., the Debye correlation function
$g^D(\rr,\rr')$.  We will refer to this Ansatz as ``Debye dynamics''. 

Maurits \etal \cite{EPD1997} proposed a particularly smart approximation to
Eq.\ (\ref{eq:onsager_chain}). They assumed that $\LL_\ab(\rr,\rr')$ is
translationally symmetric (more specifically, they postulated $\nabla
\LL_\ab(\rr,\rr') \simeq-\nabla'\LL_\ab(\rr,\rr')$). In this approximation, the
dynamical evolution equation (\ref{eq:onsager_general}) can be rewritten as a
local evolution equation for ``external potentials'' $\omega_\ga$ that are
conjugate to the density fields. Therefore, the resulting scheme is commonly
referred to as ``external potential dynamics'' (EPD). EPD calculations are very
efficient and run much faster than calculations based on other DDF schemes.
Reister \etal \cite{EPD_layer, EPD_spinodal} have compared EPD studies of
spinodal decomposition and layer formation with dynamic Monte Carlo (MC)
simulations of a coarse-grained particle-based model, and found EPD to be
superior to the corresponding local dynamcis scheme.  EPD calculations have
been used to study structure formation in blends and solutions
\cite{EPD1997,EPD_vesicle, spinodal_critical, EPD_spinodal, EPD_layer,
EPD_micelle, interface_poly, EPD_fields, spinodal_critical, EPD_composites, EPD_T},
and the approach has also been extended to include hydrodynamics \cite{heuser}.  

Other DDF schemes have focussed on strongly entangled polymer systems, where
polymers basically reptate along a tube \cite{EPD1997, Kawasaki1, Kawasaki2,
Harden, Shima}. In the present work, we will focus on the Rouse regime.

All DDF models quoted above have been constructed heuristically, and apart from
the work by Reister \etal \cite{EPD_layer, EPD_spinodal}, validations against
more fine-grained simulations are scarce. A systematic comparison with
particle-based models is therefore clearly desirable. The purpose of the
present work is to carry out such a comparison, using as a reference
``fine-grained'' system a closely related particle-based model essentially the
same monomer interactions and Rouse-type dynamics. To this end, we have
performed diffusive Brownian dynamics (BD) simulations of two classes of
inhomogeneous polymer systems: Phase separating AB polymer blends, and melts of
microphase separating A:B diblock copolymers. The Hamiltonian is expressed in
terms of local monomer densities, such that it can be directly related to a
density functional. In previous work \cite{brush_switch}, we have
compared the static properties of such particle-based models with those
obtained from static density functional theory (or self-consistent field
theory, SCF) and established the conditions under which interaction parameters
can be transferred from one model to the other without further renormalization.
Therefore, a direct comparison becomes possible.  

Methods to numerically integrate the DDF Eq.~(\ref{eq:onsager_general}) with local
dynamics Eq.~(\ref{eq:onsager_local}), Debye dynamics or EPD are well-established
and relatively straightforward \cite{fluctuation_dynamics}. However, the
implementation of DDF calculations with the full chain dynamics mobility matrix
$\LCD$ Eq.~(\ref{eq:onsager_chain}) is not trivial. One contribution of the present
work is to present an algorithm that allows us to integrate
Eq.~(\ref{eq:onsager_general}) with Eq.~(\ref{eq:onsager_chain}) directly, without
further approximations. DDF calculations based on this method can then be used
as reference to evaluate the effect of the Debye and EPD approximation. 

Specifically, we find that the EPD approximation may produce significant
artefacts in the interface broadening problem if the interfaces are sharp.
Furthermore, we find that neither the local nor the nonlocal DDF models
discussed above can capture the kinetics of microphase separation at a
quantitative level. This is most likely due to the fact that the ordering
process is driven both by local chain rearrangements, and global chain
displacements. To account for this situation, we propose a mixed local/nonlocal
coupling model, which combines global diffusion on large length scales with
local monomer motion on small length scales, inside the chain. With this mixed
scheme, reasonable agreement between the BD simulations and the DDF
calculations can be achieved.

The remainder of the paper is organized as follows: We first describe the BD
simulation model and method (Sec.\ \ref{sec:BD}) and then introduce the
different DDF schemes in more detail (Sec.\ \ref{sec:DDF}). Additional
information is given in the appendices \ref{appendix:BD} and
\ref{appendix:field_dynamics}. Furthermore, we develop in Appendix
\ref{appendix:field_dynamics} our new scheme to integrate the chain dynamic
equations. For completeness in Appendix \ref{appendix:fluctuations} we briefly describe the fluctuating DDFT schemes.
The comparison of DDF calculations with BD simulations is presented
in Sec.\ \ref{sec:results}.  We summarize and conclude in Sec.\
\ref{sec:summary}.

\section{Brownian dynamic simulations}\label{sec:BD}

We first briefly describe the BD simulation model which we use in the
reference ``fine-grained'' simulations.  To facilitate a systematic
comparison of {\em dynamic} properties with DDF calculations, we choose a type
of model whose {\em static} equilibrium properties are known to be well
reproduced by density functional theory without much parameter adjustments.
Specifically, we use Edwards-type models where the interactions are expressed
in terms of local densities \cite{Edwards, brush1994, hybrid_solution}.


We consider $n_\mr{c}$ Rouse polymers of length $N$ with mobility $D_{\mr
c}=D_{\mr o}/N$ in a box of volume $V=L_\mr x\cdot L_\mr y\cdot L_\mr z$ with
periodic boundary conditions. In the following and throughout this paper,
lengths will be given in units of the radius of gyration, $R_g$, energies in
units of the thermal energy, $k_{_B} T$, and times in units of $t_0 = R^2_\mr
g/D_{\mr c}$. 

The Hamiltonian of the system is composed of two parts,  $H=H_0+H_\mr I$, where
$H_0$ represents the contribution from the chain connectivity
\begin{equation}\label{H0_polymer}
H_0=\frac{N}{4}\sum_\mr{m=1}^{n_c} \sum_\mr{j=1}^N \Big(\mb R_\mr{m,j}-\mb R_\mr{m,j-1}\Big)^2
\end{equation}
($\mb R_\mr{m,j}$ is the position of the j-th bead of m-th chain) , and the
interaction part $H_\mr I$ is written as
\begin{equation}\label{HI_polymer}
 H_\mr I = \frac{n_\mr c\chi N}{V}\int d\mb r\hat\phi_\mr A\hat\phi_\mr B
    +\frac{n_\mr c\kappa N}{V}\int d\mb r(\hat\phi_\mr A+\hat\phi_\mr B-1)^2
\end{equation}
with the Flory-Huggins interaction parameter $\chi$ and the compressibility
(Helfand) parameter $\kappa$.  Here
$\hat\phi_\ga=\frac{1}{\rho_0}\sum_\mr{mj}\delta(\rr-\mb R_\mr{mj})
\delta_{\tau_\mr{mj} \ga}$ is the normalized microscopic density of species
$\ga = A,B$, which depends on the sequence $\tau_\mr{mj}$ of monomers $A,B$ on
the mth chain ($\tau_\mr{mj}=A,B$). It is normalized with respect to the mean
monomer density $\rho_0=n_\mr{c} N/V$. The equation of motion for a BD bead is
controlled by a deterministic conservative force derived from the Hamiltonian
and a random force, 
\begin{equation}
\frac{\ud \mb R_\mr{m,j}}{\ud t}
  =-D_\mr{o} \frac{\partial H}{\partial\mb R_\mr{m,j}}
   +\sqrt{2 D_\mr{o}} \: \mb f_\mr{m,j}
\end{equation}
with $D_\mr{o} = \frac{1}{N}$ (in units of $R_g^2/t_0$). The random force $\mb
f_\mathrm{mj}$ is Gaussian distributed with zero mean and variance $\langle
f_\mr{mjI}(t) f_\mr{nkJ}(t')\rangle =
\delta_\mr{mn}\delta_\mr{jk}\delta_\mr{IJ}\delta(t-t')$ where I,J denote the
Cartesian components. The derivative of the Hamiltonian with respect to the
bead position can be evaluated directly, giving 
\begin{equation}
\frac{\partial H}{\partial\mb R_\mr{m,j}}=\frac{N}{2}\big(2\mb R_\mr{m,j}-\mb R_\mr{m,j+1}-\mb R_\mr{m,j-1}\big)
  +\frac{1}{N}\frac{\partial u_{\tau_\mr{mj}}}{\partial\mb R_\mr{m,j}}
\end{equation}
where $u_\ga$ is given by
\begin{equation}
  u_\ga = \chi N \sum_\gb \hat\phi_\gb  (1-\delta_\ab) 
    + \kappa N\big[\hat\phi_\mr A+\hat\phi_\mr B-1\big].
\end{equation}

In practice, the microscopic densities are evaluated on a grid using an
assignment scheme, which assigns densities to mesh points based on the bead
positions. Here we use a first order  CIC scheme \cite{CIC}, where
beads contribute to the densities of the eight closest mesh points. Details are
given in Appendix \ref{appendix:BD}. The BD equation is a stochastic
differential equation, and it is integrated using the explicit Euler-Maruyama
method. Other schemes with higher order accuracy are conceivable
\cite{stochastic_numerical}.

The present BD scheme has similarities to the ``hybrid particle-field'' BD
scheme proposed by Ganesan and coworkers \cite{ganesan1,interface_SCFBD}.
However, we wish to stress that our simulations here do not involve a
mean-field approximation. We carry out ``true'' BD simulations of a particle
system with the well-defined Hamiltonian $H$. In the particle-field scheme of
Ganesan et al, and in related schemes such as the SCMF scheme by M\"uller and
coworkers \cite{scmf1,scmf2} and ``hybrid particle-field molecular dynamics''
by Milano and coworkers \cite{hybrid_MD, Sevink}, monomers move in ``mean
potentials fields'' $\omega_\alpha$. These are separate variables that evolve
according to their own dynamics, which is deliberately chosen slower than the
bead dynamics. In our simulations, these ``mean potentials'' are replaced by
the actual instantaneous interactions $u_\alpha$ determined from the
Hamiltonian $H_\mr{I}$.


Since the interactions are defined in terms of densities, which are evaluated
on a grid, the size of the grid cells is an important component of Edwards-type
models. In fact, it determines the range of nonbonded interactions. The system
can only undergo (micro)phase separation if the number of interacting
particles in a cell is sufficiently high. In the present simulations, we use
grid sizes in the range of 0.2--0.25 $R_g$ (see below), and the number of
particles in a cell is about 50.  With these parameters, lattice artefacts were
found to be negligibly small. For a detailed analysis of discretization
effects, we refer to the references \cite{particle_mesh, brush_switch}.

The BD scheme presented above can be viewed as an approximate dynamic
Monte Carlo (MC) scheme for studying systems with Edwards-type density-based
interactions $H_\mr{I}$.  The idea to use such Hamiltonians in MC simulations
was first put forward by Laradji et al \cite{brush1994}, and later applied in
studies of a variety of inhomogeneous polymer systems\cite{EdwardsType2, scmf2,
particle_mesh, hybrid, hybrid_solution, brush_switch}.  They have the advantage
of being computationally efficient, since the explicit evaluation of the pair
interaction, which is often the most time consuming part in a simulation, is
circumvented.  The main reason why we choose this class of systems as BD
reference systems is that monomer interactions in the BD model and the
field-based dynamic models are described by the same expression, the
Hamiltonian $H_\mr{I}$, hence the results can be compared directly. In fact,
Hamiltonians such as $H_\mr{I}$ are typically taken as a starting point to
derive field theoretic descriptions of polymer systems and DDF theories such as
those described in the next section. Therefore, a comparison of DDF predictions
with explicit simulations of the same model should be particularly meaningful.


\section{Dynamic density functional approaches}\label{sec:DDF}

Within density functional theory or (equivalently) self-consistent field
theory, the model systems introduced in the previous section are described by a
free energy functional of the form \cite{review_scf} $\mc{F}[\{\rho_\ga\}] =
\frac{n_\mr c}{V} F[\{\phi_\ga\}]$ with
\begin{eqnarray}
\label{eq:scf_free_energy}
F &=& \int \ud \rr \Big[ \chi N \phi_A \phi_B 
  + \kappa N (\phi_A+\phi_B-1)^2 \Big]
\nonumber \\
&& - \: \sum_{\ga=A,B} \int \ud \rr \: \phi_\ga \omega_\ga 
     - V \sum_\gamma \frac{n_\gamma}{n_\mr{c}} \ln \mc{Q}_\gamma.
\end{eqnarray}
Here $\phi_\ga(\rr)$ are the mean normalized density fields ($\phi_\ga(\rr) =
\rho_\ga(\rr)/\rho_0$ with $\rho_0=n_\mr{c}N/V$), $\omega_\ga(\rr)$ are
conjugated ``potential'' fields, $n_\gamma$ is the number of chains of type
$\gamma$ and $\mc{Q}_\gamma$ the corresponding single chain partition function.
In homopolymer blends, we have two chain types corresponding to chains $A$ and
$B$, whereas copolymer melts contain only one type of chain. 

The density fields $\{ \phi_\ga(\rr)\}$ and the $\mc{Q}_\gamma$ can be
calculated from the potential fields $\omega_\ga(\rr)$ via the following
explicit procedure: One parametrizes the contour of chains with a
continuous variable $s \in [0:1]$, such that the mth monomer corresponds to
$s=\mr{m}/N$, and describe the monomer sequence on chains of type $\gamma$ by a
function $\tau_\gamma(s)$ ($\tau = A,B$).  Furthermore, we introduce partial
partition functions $q_\gamma(\rr,s)$ and $q'_\gamma(\rr,s)$ that satisfy the
diffusion equation
\begin{eqnarray}
\partial_s q_\gamma(\rr,s) 
  &=& \Delta q_\gamma - \omega_{\tau_\gamma(s)}(\rr) q_\gamma
\nonumber \\
\partial_s q'_\gamma(\rr,s) 
  &=& \Delta q'_\gamma - \omega_{\tau_\gamma(1-s)}(\rr) q'_\gamma
\label{eq:scf_qq}
\end{eqnarray}
with initial condition $q_\gamma(\rr,0)=q'_\gamma(\rr,0)\equiv 1$. Then, the
single chain partition functions are given by $\mc{Q}_\gamma=\int \ud \rr \:
q_\gamma(\rr,1)$, and the density fields $\phi_\ga(\rr)$ can be calculated via
\begin{equation}
\label {eq:scf_phi}
\phi_\ga(\rr) = \frac{1}{\rho_0} \sum_\gamma \frac{n_\gamma N}{\mc{Q}_\gamma} 
   \int_0^1 \ud s \: q_\gamma(\rr,s) \: q'_\gamma(\rr,1-s) 
   \delta_{\ga,\tau_\gamma(s)}.
\end{equation}

The inverse determination of $\{\phi_\ga\}$ as a function of $\{\omega_ \ga\}$ cannot be
done explicitly, it requires the use of iteration techniques. More details and
the derivations of Eqs.\ (\ref{eq:scf_free_energy}-\ref{eq:scf_phi}) can be
found, e.g., in Ref.\ \cite{fluctuation_dynamics, review_scf, review2_scf}.

The free energy functional, Eq.\ (\ref{eq:scf_free_energy}), enters the DDF
equation (\ref{eq:onsager_general}) through the ``local chemical potential'',
$\mu_\ga = \delta \mc{F}/\delta \rho_\alpha$. Rewriting
(\ref{eq:onsager_general}) in terms of renormalized densities $\phi_\ga$ and
using $D_\mr{o}=N^{-1}$, we obtain
\begin{equation}
\label{eq:ddf_general}
 \partial_t \phi_\ga = \nabla \int \ud \rr' \tLL_\ab(\rr,\rr') 
     \: \nabla' \tilde{\mu}_\gb(\rr')
\end{equation}
with $\tilde{\mu}_\ga = N \mu_\ga = \delta F[\{\phi\}]/\delta \phi_\ga$ and
$\tLL_\ab = (\rho_0 N)^{-1}\LL_\ab$. Specifically, one gets
\begin{equation}
 \tilde{\mu}_\ga = \chi N \sum_\gb \phi_\gb(1-\delta_\ab) 
      + 2 \kappa N(\phi_A+\phi_B - 1) - \omega_\ga,
\end{equation} 
and the dynamic coupling types discussed in the introduction correspond to
the rescaled Onsager matrices 
\begin{eqnarray}
\label{eq:ddf_local}
\mbox{Local dynamics:} \; \tLLD_\ab(\rr,\rr') 
   &=& \phi_\ga(\rr) \: \delta_\ab \: \delta(\rr-\rr') \\
\label{eq:ddf_chain}
\mbox{Chain dynamics:} \; \tLCD_\ab(\rr,\rr') 
   &=& - \delta\phi_\ga(\rr)/\delta \omega_\gb(\rr') \\
\label{eq:ddf_debye}
\mbox{Debye dynamics:} \; \tLDB_\ab(\rr,\rr') 
   &=& \sum_\gamma \bar{\phi}_\gamma g_\ab^{D,\gamma}(\rr-\rr')
\end{eqnarray}
(in simulation units), where $\bar{\phi}_\gamma$ is the (normalized) mean
density of monomers in chains of type $\gamma$, and
$g_\ab^{D,\gamma}(\rr-\rr')$ the Debye correlation function of these chains
\cite{Doi_book}. Specific expressions for $g^D_\ab$ for the chains considered
in the applications in this work are given in Appendix
\ref{appendix:field_dynamics}.  Finally, the EPD dynamical equation can be
written as
\begin{equation}
\label{eq:ddf_epd}
\mbox{EPD dynamics:} \quad \partial_t \omega_\ga = - \nabla^2 \tilde{\mu}_\ga.
\end{equation}

The dynamical equations (\ref{eq:ddf_local}-\ref{eq:ddf_epd}) are discussed in
more detail in Appendix \ref{appendix:field_dynamics}. Furthermore, we
present our new numerical scheme that allows us to integrate the DDF
equations with the Onsager matrix $\tLCD_\ab$ without further approximations. 

As we will see below in Sec.\ \ref{sec:results_microphase}, neither the local
coupling scheme Eq.~(\ref{eq:ddf_local}) nor the nonlocal schemes
Eqs.~(\ref{eq:ddf_chain} - \ref{eq:ddf_epd}) provide a satisfactory description
of the kinetics of microphase separation. This is because the local dynamics
scheme disregards chain connectivity, whereas the chain dynamics scheme
overemphasizes it. In reality, monomers of a chain move together on larger
scales, but they are free to rearrange locally on shorter scales below $R_g$. 

To account for this situation at least at an approximate level, we propose a
heuristic scheme that combines local dynamics on small scales with nonolcal
dynamics on large scales: The idea is to interpolate the Onsager functions such
that they assume the form of local dynamics on scales below $R_g$ and the form
of nonlocal dynamics on large scales of order $R_g$ and larger. To this end,
we introduce a filter function
\begin{equation}
\label{eq:filter}
\Gamma(\rr) = (2 \pi \sigma^2)^{-3/2} \: \exp(-r^2/2 \sigma^2),
\end{equation}
where $\sigma < R_g$ is a tunable parameter which determines the length scale
of crossover between local and nonlocal dynamics. The filter function is used
to separate the (rescaled) thermodynamic force acting on monomers $\ga$,
$\ff_\ga = - \nabla \tilde{\mu}_\ga(\rr)$, into a ``coarse-grained'' part that
governs the global behavior,
\begin{equation}
\ff^{CG}_\ga(\rr) = \int \ud \rr' \Gamma(\rr-\rr') \: \ff_\ga(\rr'), 
\end{equation}
and a remaining ``fine-grained'' part that drives local rearrangements
\begin{equation}
\ff_\ga^{FG}(\rr) = \ff_\ga(\rr) - \ff^{CG}_\ga(\rr)
  = \int \ud \rr' \Big[ \delta(\rr-\rr') - \Gamma(\rr-\rr')\Big] \ff_\ga(\rr').
\end{equation}
The interpolated DDF equation for $\phi_\ga$ is then given by
\begin{eqnarray}
  \partial_t \phi_\ga &=& - \nabla \sum_\gb \int \ud \rr' 
  \Big[ \tLNL_\ab(\rr,\rr') \ff^{CG}_\gb(\rr') 
\nonumber \\
  && \quad + \: \tLLD_\ab(\rr,\rr') \ff^{FG}_\gb(\rr') 
\Big],
\label{eq:ddf_mixed}
\end{eqnarray}
where $\tLNL$ refers to one of the nonlocal Onsager matrices discussed above
(Eq.\ (\ref{eq:ddf_chain}-\ref{eq:ddf_epd})). In the present work, we use Debye
coupling (\ref{eq:ddf_debye}). 

We close this section with a remark on thermal fluctuations. In the present study, they are disregarded, i.e., we compare the BD simulation 
results with mean-field DDF calculations. A measure for the fluctuation effect is the so called Ginzburg parameter $C^{-1}$ \cite{polymer_book}, i.e. the amplitude of thermal noise scales with the inverse Ginzburg parameter
$C^{-1} = k_B T \: V / n_\mr c R_g^3$. (Here we have reinserted the energy 
and length units). In the BD simulations described below, this parameter 
is of order $C^{-1} \sim 0.01$ or less, hence fluctuations are not expected
to be important. However, for dilute system, where $C^{-1}$ is large, fluctuation effect should be taken into account. In Appendix \ref{appendix:fluctuations} we briefly describe how to incorporate fluctuations into the DDFT equations.

All these DDFT equations are integrated numerically.  Spatial derivatives
are always evaluated in Fourier space using fast Fourier transformations. Apart
from that, the local dynamics, chain dynamics and mixed dynamics equations are
integrated in real space using the explicit Euler scheme
with time steps as specified below in sec.~\ref{sec:results}.
Simulations with different time steps were also performed in
selected cases to ensure that the results do not depend on the time step. The
Debye dynamics and EPD equations are integrated in Fourier space using a
semi-implicit scheme proposed by Fredrickson and coworkers
\cite{polymer_book,semi_implicit}.
Here, we also checked explicit schemes and different time steps, and found that
the results were the same.

\section{Results and discussion} \label{sec:results}

We will now test the DDF schemes described in the previous section by comparing
their predictions with explicit BD simulations (see Sec.\ \ref{sec:BD}). We
begin with studying the time evolution of an AB interface in a blend of
incompatible A/B homopolymers after a sudden drop of the interaction parameter
$\chi$. Then, we study the dynamics of microphase separation in initially
disordered A:B diblock copolymer melts.

\subsection{Interface evolution in an A/B polymer blend}
\label{sec:results_interface}

To study the chain interdiffusion, we consider a compressible polymer blend
composed of $n_\mr A$ A homopolymer and $n_\mr B$ B homopolymers. 
All chains have the same length $N$. The reference density is
defined as $\rho_0=nN/V$, where $n=n_\mr A+n_\mr B$ is the total number of
chains, and in the present study we set $n_\mr A=n_\mr B$. 

We restrict ourselves to systems where polymers are uniformly distributed along
the $x$ and $y$ directions, and interfaces appear only along the $z$ direction.
Due to the periodic boundary conditions, the system then contains two
interfaces. In the BD scheme, we perform three-dimensional simulations, and
calculate the one-dimensional density profiles by averaging the
three-dimensional density over the $x$ and $y$ coordinates. In the field-based
DDF schemes, we perform one-dimensional calculations. The two type of polymers
in the system are completely symmetric, i.e., they have the same properties. In
the following discussion, we therefore only focus on the A chains.

We should note that the interface in the three dimensional BD system is subject
to capillary wave fluctuations and broadening, which are neglected in our
mean-field DDF calculations. In three dimensions, the capillary wave 
broadening grows logarithmically on the lateral system size. Previous
work \cite{Werner} has shown that SCF predictions for interfaces in polymer
blends are in good agreement with simulation results on length scales 
comparable to the interfacial width. In the BD simulations, we therefore use
simulation boxes which are small in the ($x,y$) direction, $L_\mr x = L_\mr y =
2$. The system size in the $z$ direction is chosen $L_\mr z=16$ both in the BD
simulations and in the DDF calculations, and the number of grid points is
$n_\mr z=64, n_\mr x = n_\mr y = 8$.  The systems contain 10.000 chains of
length $N=20$. Thus the average number of beads in each cell is about 50. 
The compressibility parameter is set to $\kappa N=10$. 

The time steps depend on the method. In the BD simulations and the DDF
calculations based on local dynamics, chain dynamics, and Debye dynamics, we
use $\Delta t=10^{-4} t_0$, and in EPD dynamics, we use $\Delta t=10^{-3} t_0$.
We verified in all cases that the results do not depend on the time step. In
EPD, one could choose even larger time steps. In the other DDF schemes,
densities sometimes became negative if the time steps were too large. 
The spatial discretization in the DDF calculations is $\Delta z = 0.25 R_g$.

We first study the process of interfacial broadening. The initial density is
constructed as a sharp ``physical" density profile which is obtained by
equilibrating the interfaces at $\chi N=8$. At time $t=0$, $\chi N$ suddenly
drops to a smaller value. As a result, the interfaces broaden, the density
profiles become more diffuse, until they finally reach a new equilibrium state.
We monitor the density profiles at the interface as a function of time. The
width of the interface is simply defined as the inverse of the maximum slope of
the density profiles.  We do not renormalize this quantity with respect to the
``bulk densities'' (as is usually done), because the latter also change in
response to the change of $\chi N$ and are not always well-defined during the
interdiffusion process.

\begin{figure}[htb]
  \centering
    \includegraphics[angle=0, width=7.0cm]{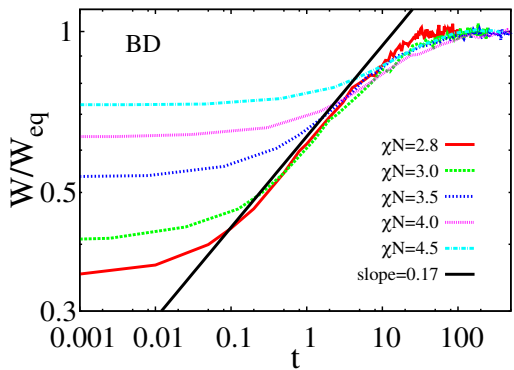}
  \caption{Evolution of the rescaled width $W$ of A/B interfaces in
incompatible homopolymer blends as a function of time $t$ (in units of $t_0$)
in double logarithmic representation, as obtained from BD simulations for
different final incompatibility parameters $\chi N$ as indicated. The inital
state is an equilibrated interface at $\chi N = 8$. The data for $W$ are
rescaled with the equilibrium width of the final state. The solid black line
indicates an apparent power law $W(t)\propto t^a$ with $a\approx0.17$.
The magnitude of the errorbars here is comparable to that shown in Fig.(\ref{fig:width_chi}).}
  \label{fig:width} 
\end{figure}

Fig.\ \ref{fig:width} shows the evolution of the interfacial width $W$ as a
function of time after such a sudden jump from $\chi N=8$ to different smaller
values of $\chi N$, as obtained in the BD simulations. The data for $W$ are
rescaled by the corresponding final, equilibrated value. Except for early
times, the curves for different $\chi N$ parameters collapse onto a single
master curve, which follows an apparent scaling relation $W(t)\propto t^a$ at
intermediate times \cite{interface_FH} with exponent $a \approx 0.17$.    

\begin{figure}[htb]
  \centering
  \subfigure[]{
    \label{fig:2a} 
    \includegraphics[angle=0, width=7.0cm]{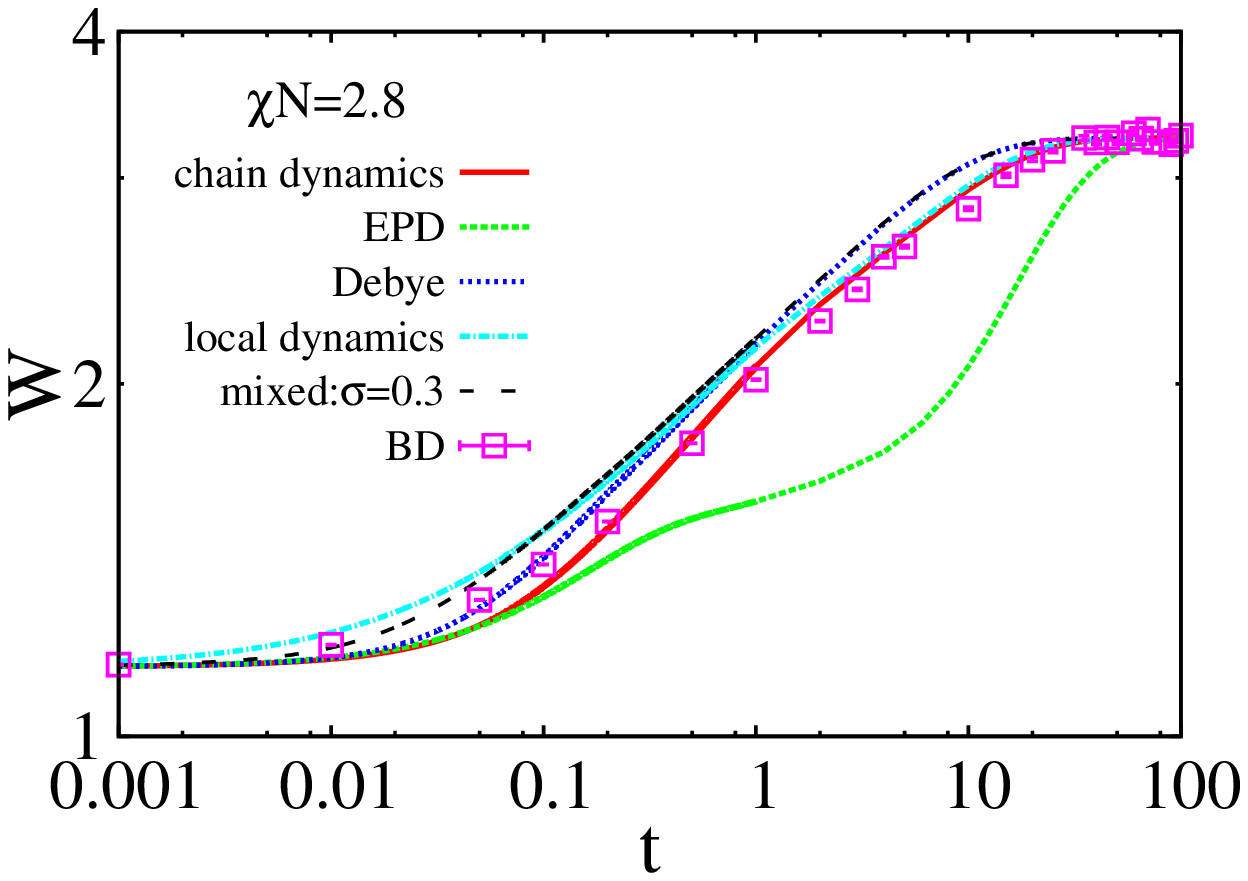}}
  \subfigure[]{
    \label{fig:2b} 
    \includegraphics[angle=0, width=7.0cm]{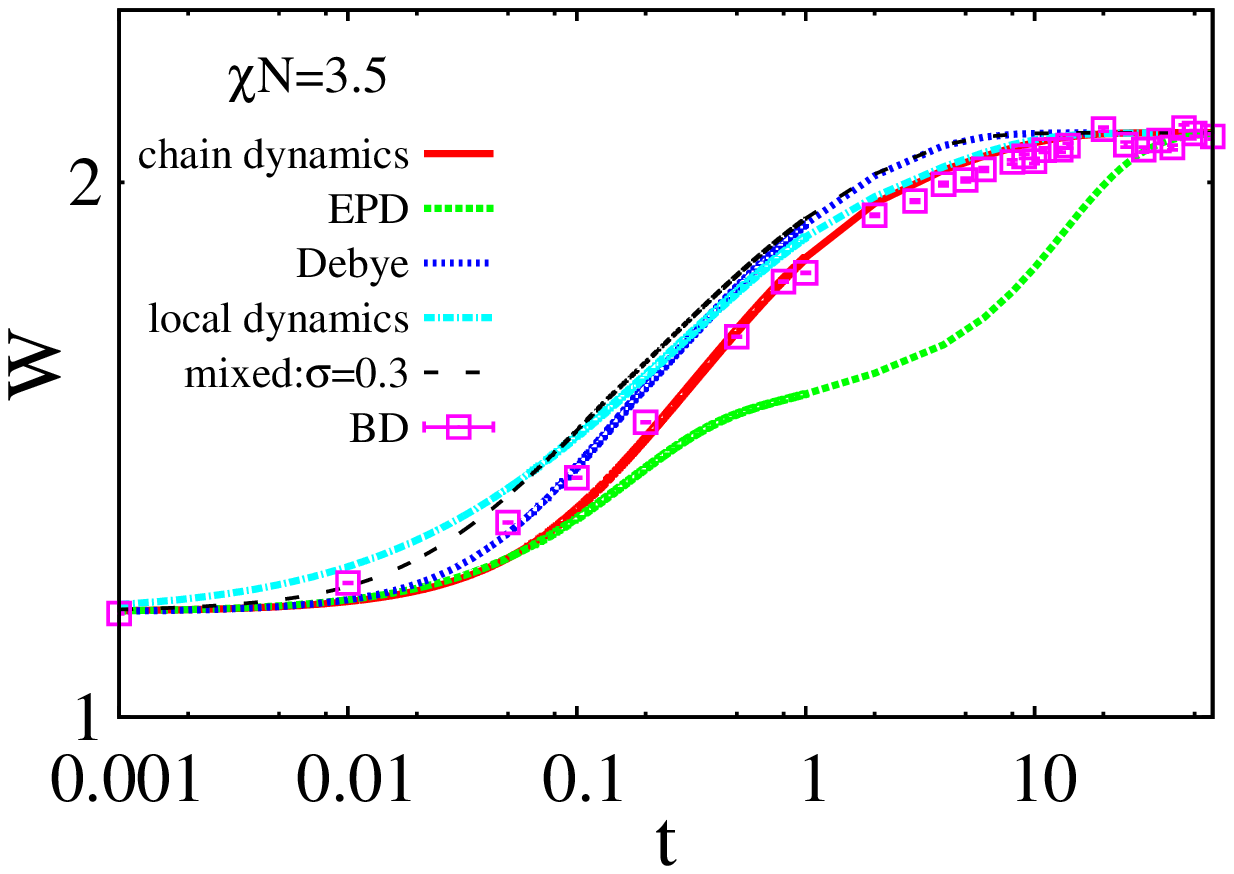}}
  \caption{Interfacial width (in units of $R_g$) as a function of time (in
units of $t_0$) in double logarithmic representation, as obtained by
different DDF models as indicated (chain dynamics, EPD, Debye dynamics,
local dynamics, and mixed dynamics with $\sigma=0.3$), compared to the
BD simulation results (symbols), for $\chi N=2.8$ (a) and $\chi N=3.5$ (b).
The initial configuration corresponds to an equilibrated interface at 
$\chi N=8$. The magnitude of errorbars is smaller than the size of the symbols.
}
  \label{fig:width_chi} 
\end{figure}

Figure \ref{fig:width_chi} compares the predictions for the broadening of the
interfacial width from the different DDF models discussed in Sec.\
\ref{sec:DDF} with the BD results for two examples of $\chi N$ jumps.  The
predictions of the chain dynamics model agree well with the BD simulation
results over almost the whole time interval under consideration (except for
very early times $t < 0.2 t_0$). The predictions from local dynamics
calculations differ slightly, but noticeably from the BD results at early
times, and approach them at $t \simeq 4 t_0$.  These results indicate that
interfacial broadening is mainly driven by chain diffusion, while the effect of
segmental dynamics is minor.  

The Debye dynamics calculations reproduce the BD results well at early times,
but show small deviations at later times. This is presumably due to the
underlying weak inhomogeneity assumption, which is clearly questionable in the
presence of sharp interfaces. The mixed model, which is partly based on the
Debye model, does not improve on this problem.  Overall, however, all DDF
predictions discussed so far are acceptable compared to the reference BD
simulations; the deviations are quite small. When plotting the DDF
predictions for different final $\chi N$ in a similar fashion as in Fig.\
\ref{fig:width}, the data collapse onto a master curve and exhibit the same
apparent power law behavior at intermediate times than the BD data, $W(t) \propto t^a$
with $a\approx 0.17$ (data not shown). 

\begin{figure}[htb]
  \centering
      \subfigure[]{
    \label{fig:3a} 
    \includegraphics[angle=0, width=7.0cm]{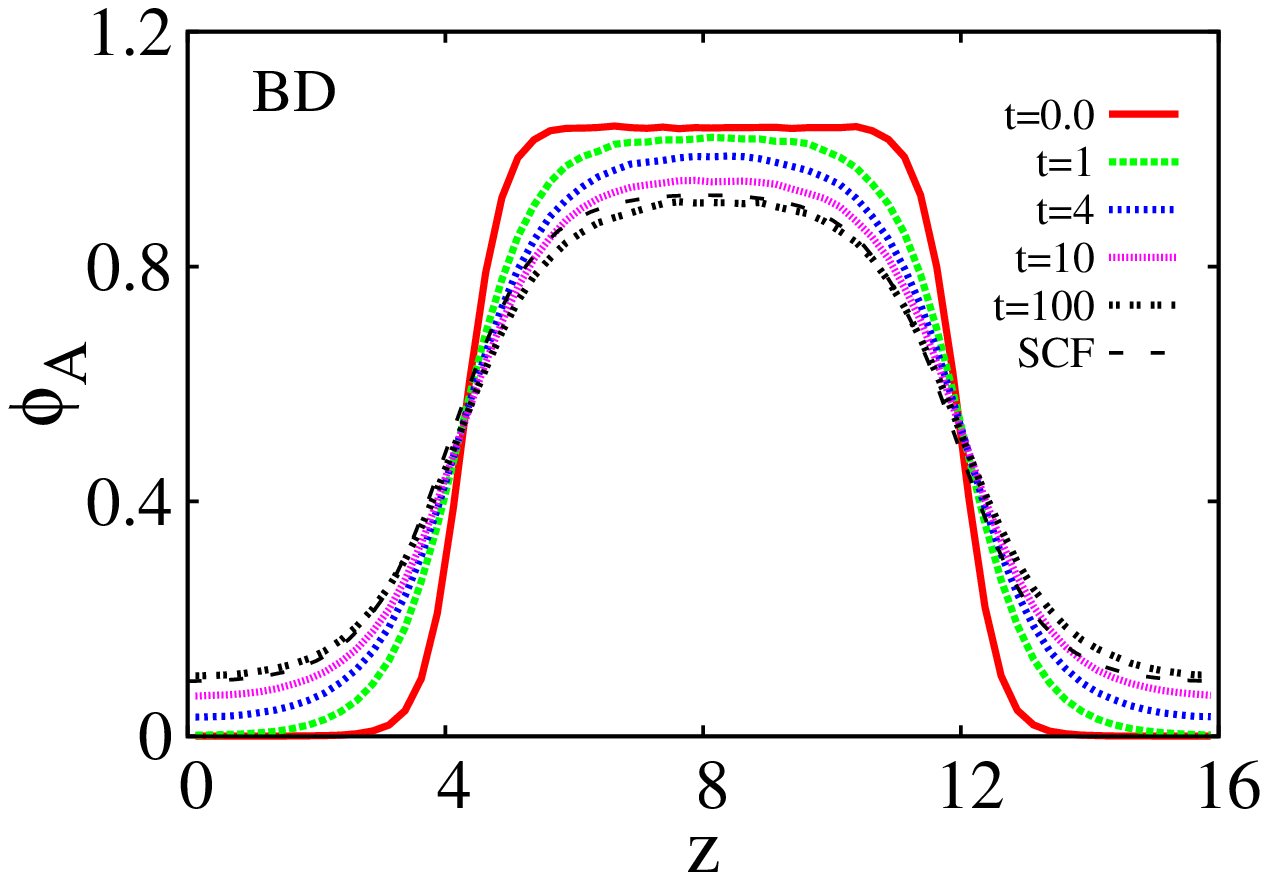}}
  \subfigure[]{
    \label{fig:3b} 
    \includegraphics[angle=0, width=7.0cm]{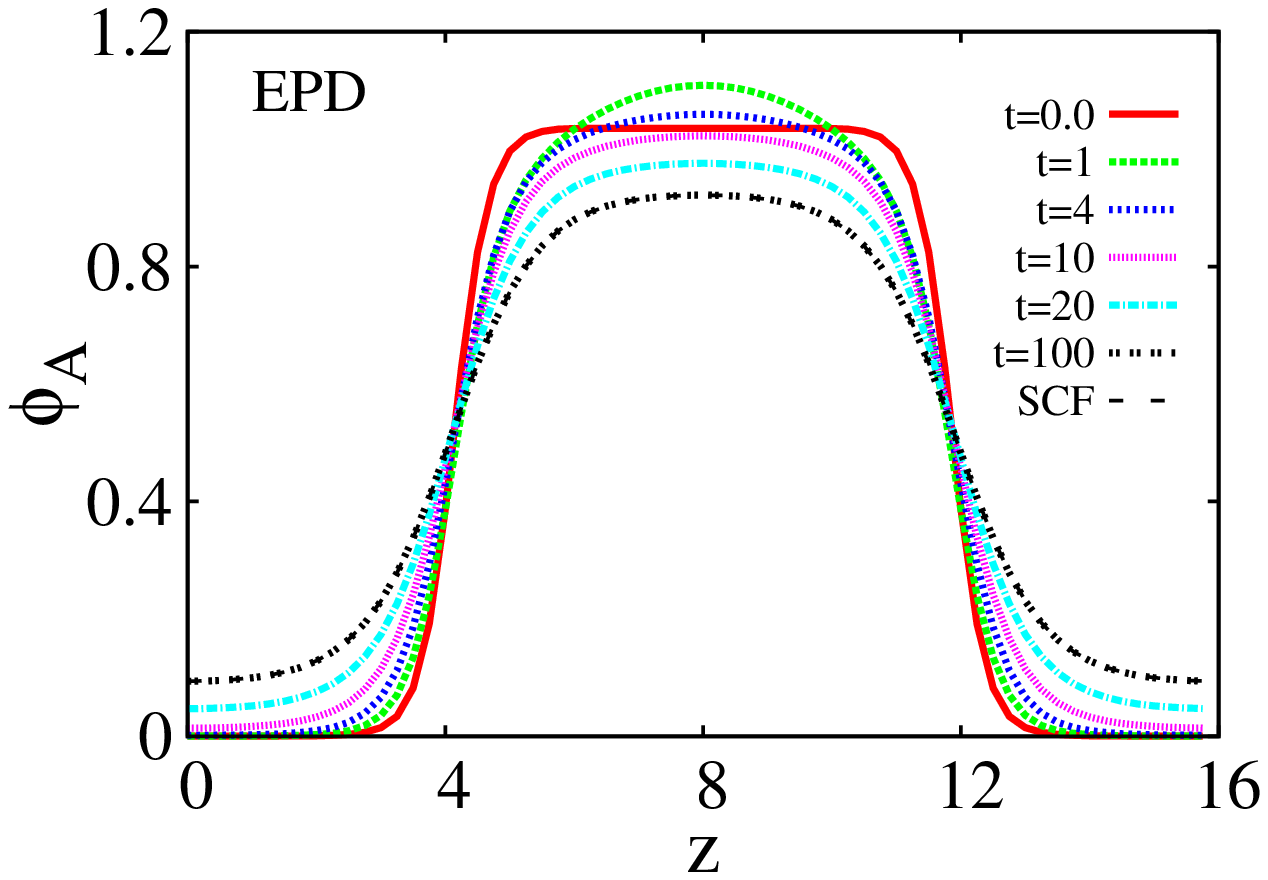}}
  \caption{Density profiles of component A obtained in BD simulations (a), and
EPD calculations (b) for different times $t$ (in units of $t_0$) as indicated.
The system is initialized at $\chi N = 8$ and then suddenly set to $\chi N =
2.8$. The dashed line shows the equilibrium density profile at $\chi N =2.8$ as
obtained from self-consistent field (SCF) theory.}
  \label{fig:density_evolution} 
\end{figure}

This is different for the EPD model. The EPD predictions for the evolution of
the interfacial width in Fig.\ \ref{fig:width_chi} differ strongly from the BD
reference data over a wide time window, $0.5 t_0 < t < 40 t_0$. The values for
the width are much too small, i.e., the interface broadening is slowed down
significantly, and the shape of the curves is very different.  To further
analyze this problem, we compare in Fig.\ \ref{fig:density_evolution}a) and b)
the evolution of the A-density profiles obtained from the EPD model with the
profiles in the reference BD system.  According to the BD simulations (Fig.\
\ref{fig:density_evolution}a), A-polymer chains gradually diffuse from the
A-polymer rich region to the A-polymer poor region. The amount of A-polymers in
the A polymer rich region decreases monotonically with time. The corresponding
curves obtained from chain dynamics, local dynamics, Debye, and mixed dynamics
calculations are very similar (data not shown). In contrast, the EPD
calculation produces a highly unusual, nonmonotonic behavior (Fig.\
\ref{fig:density_evolution}b). At intermediate times, A-chains accumulate in
the middle of the A-polymer rich slab, such that the A-density there even
exceeds the initial value. A shallow peak forms in the middle of the A-slab,
which reaches a maximum and then decreases again.  The slowdown of interfacial
broadening in Fig.\ \ref{fig:width_chi} is observed precisely in the time range
where the peak is highest.

This spurious behavior is only found in the EPD model, and not in the closely
related chain dynamics model. It must thus be an artefact of the EPD
approximation.  If the EPD assumption, $\nabla \LL_\ab(\rr,\rr') \simeq
-\nabla'\LL_\ab(\rr,\rr')$), is not valid, the EPD equations violate local mass
conservation, since they no longer have the form of a continuity equation for
the densities $\phi_\ga$. Instead, the auxiliary potentials $\omega_\ga$ are
locally conserved. Large-scale chain redistributions become possible, which are
presumably responsible for the observed artefacts.

\begin{figure}[htb]
  \centering
      \subfigure[]{
    \label{fig:4a} 
    \includegraphics[angle=0, width=7.0cm]{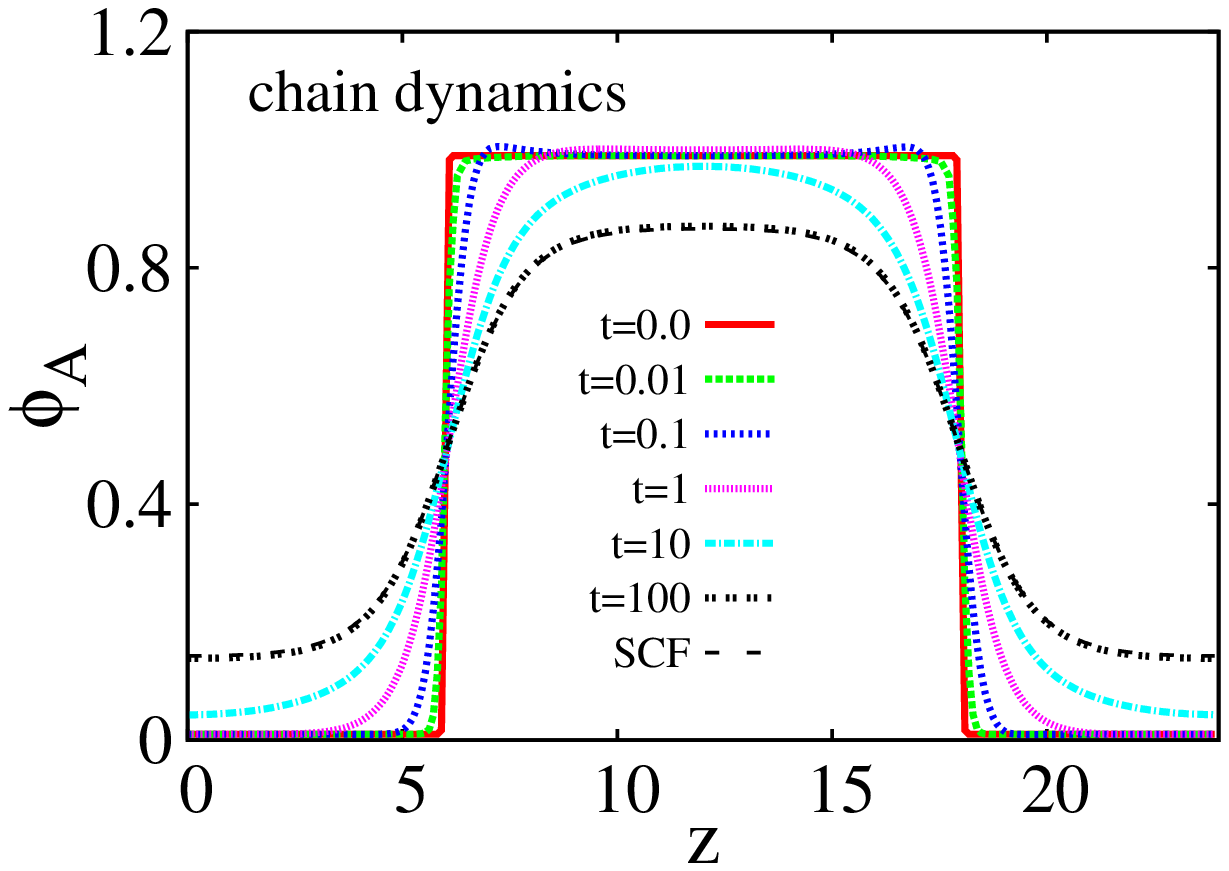}}
      \subfigure[]{
    \label{fig:4b} 
    \includegraphics[angle=0, width=7.0cm]{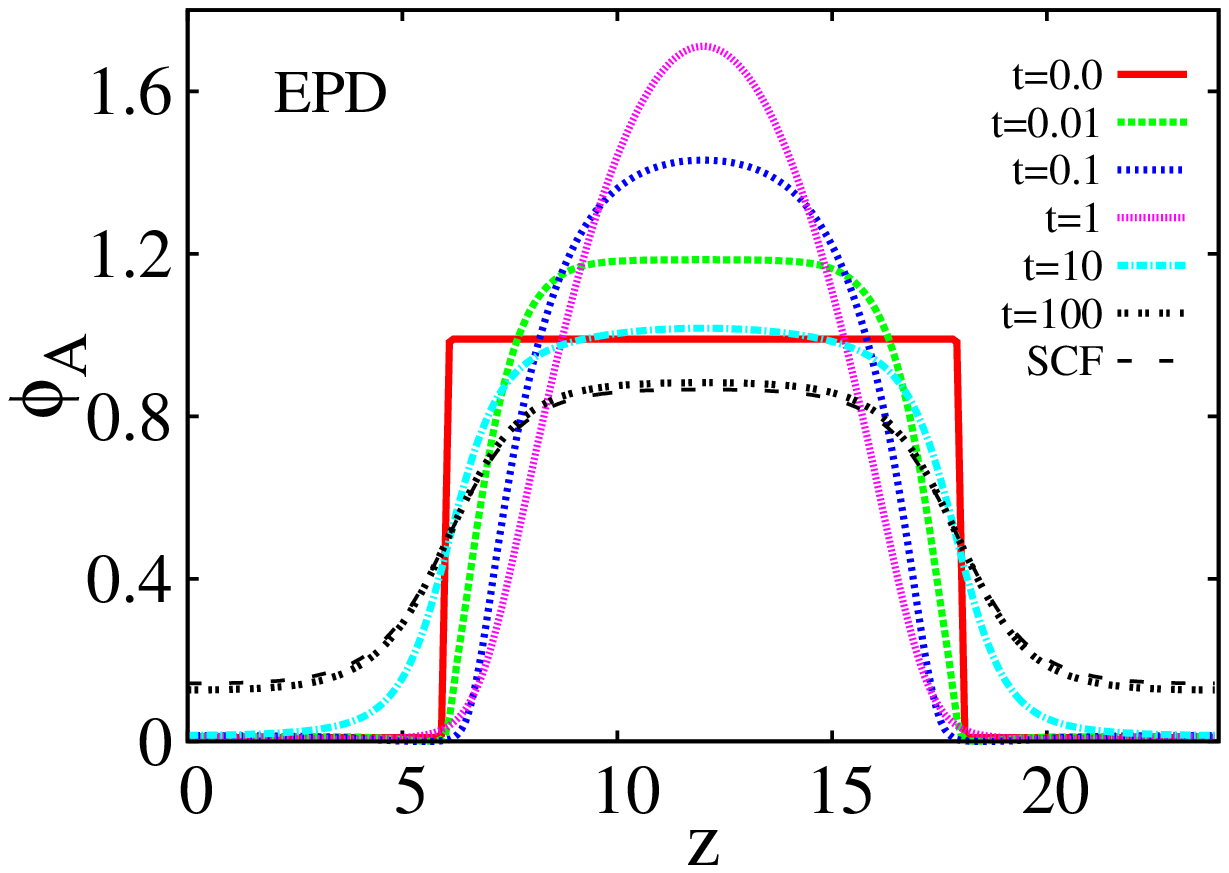}}
  \caption{Density profiles of component A obtained from chain dynamics
calculations (a), and EPD calculations (b) for different times $t$ (in units of
$t_0$) as indicated.  The system is initialized with two sharp interfaces and
then further propagated with $\chi N = 2.5$.  The dashed line indicates the
equilibrium density profile at $\chi N =2.5$ according to the SCF theory.}
  \label{fig:density_tanh} 
\end{figure}

To further illustrate this problem, we now consider an extreme case
where the initial density profile has an almost steplike configuration,
\begin{equation}
  \label{eq:tanh}
  \phi_\mr A^\mr{ini} = \frac{\phi_\mr{min}+\phi_\mr{max}}{2}
   + \frac{\phi_\mr{min}-\phi_\mr{max}}{2\tanh(\eta)}
    \tanh\Big[\eta\cos\frac{2\pi}{L_\mr z}z\Big]
\end{equation}
with $\phi_\mr{min}=0.01, \phi_\mr{max}=0.99$, and $\eta = 100$.  The initial
interfacial width is roughly $W \simeq 0.19$.  Such unphysical sharp density
profiles were also adopted in some other theoretical studies \cite{Shi_interface,interface_poly}.
Since they cannot easily be generated in particle simulations, we
study the evolution of the density profile by DDF methods only.  The system
size was  chosen $L_\mr z=24$ with spatial discretization $\Delta z = 0.09 R_g$
($n_\mr z=256$ grid points).

Fig.\ \ref{fig:density_tanh} illustrates that the EPD artefacts become even
more prominent for such sharp interfaces. The results from the chain dynamics
calculations, shown in Fig.\ \ref{fig:density_tanh}a, seem quite reasonable.
At early times, weak maxima of A-monomer density appear in the A-polymer rich
side right next to the interface, which then gradually move away and shrink. A
similar phenomenon was reported from DDF studies of interfacial broadening in
incompressible polymer blends \cite{Shi_interface,interface_poly}. Apart from
this overshooting, the evolution of the interfacial density profiles is similar
to that shown in Fig.\ \ref{fig:density_evolution}a). In contrast, the results
from EPD calculations (Fig.\ \ref{fig:density_tanh}a) are obviously erroneous.
The density of A-chains at the center of the A-slab grows by a factor of almost
2 within a time span of less than $1 t_0$, which is not compatible with a
regular chain diffusion process. 

\begin{figure}[htb]
  \centering{\includegraphics[angle=0, width=7.0cm]{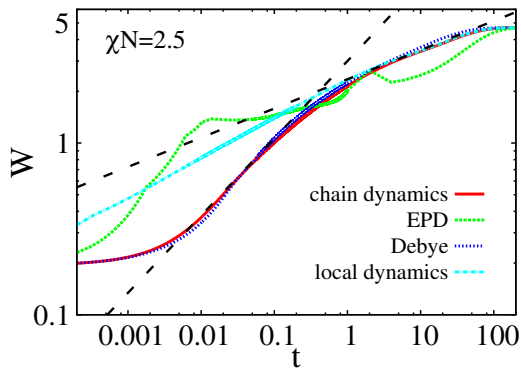}}
  \caption{Interfacial broadening with time according to different
DDF models as indicated. The initial density profile is a sharp tanh function
with width $0.19 R_g$, and the interaction parameter is $\chi N=2.5$.
Dashed lines indicate power laws $W(t) \propto t^a$ with exponents 
$a=0.45$ and $a=0.17$.}
  \label{fig:width_tanh} 
\end{figure}

Fig.\ (\ref{fig:width_tanh}) shows the evolution of the interfacial width
obtained with different DDF methods for systems that were initialized with the
sharp tanh function (\ref{eq:tanh}). The results obtained from chain dynamics
and Debye dynamics calculations are in good agreement with each other. The
curve corresponding to local dynamics differs significantly from the nonlocal
schemes at early times and approaches it at later times. Since chain diffusion
is a dominant mechanism in interfacial broadening, we expect that the nonlocal
schemes reflect the true dynamics more accurately than the local dynamics
scheme.  The curve produced by the EPD dynamics is rather erratic and has no
resemblance to the other curves.

In contrast to Fig.\ \ref{fig:width}, the curves $W(t)$ obtained with the
nonlocal schemes (chain dynamics / Debye dynamics) cannot by described by a
single scaling law $W(t) \propto t^a$.  We therefore extract two power law
exponents, characterizing the interface broadening at early and later times. At
early times, we get $a \approx 0.45$, and at late times $a \approx 0.17$.
Experimental studies have also indicated a power law growth of the interfacial
width before saturating to the final equilibrium width. The power law exponent
reported in the experiments falls between $0.25\sim0.5$ depending mainly on the
annealing temperature \cite{blend_E1,blend_E2}. Our predicted power law
exponent for early times (see Fig~\ref{fig:width_tanh} matches well with these
experimental results. On the other hand, the smaller apparent exponent $a \approx 0.17$
at later times is compatible with the apparent exponent $a \approx 0.17$ extracted
from Fig.~\ref{fig:width}. This is not surprising, as the initial interfaces in
Fig.~\ref{fig:width} are much wider, and the broadening starts to saturate much
earlier.

\begin{figure}[htb]
  \centering
  \subfigure[]{
    \label{fig:6a} 
    \includegraphics[angle=0, width=7.0cm]{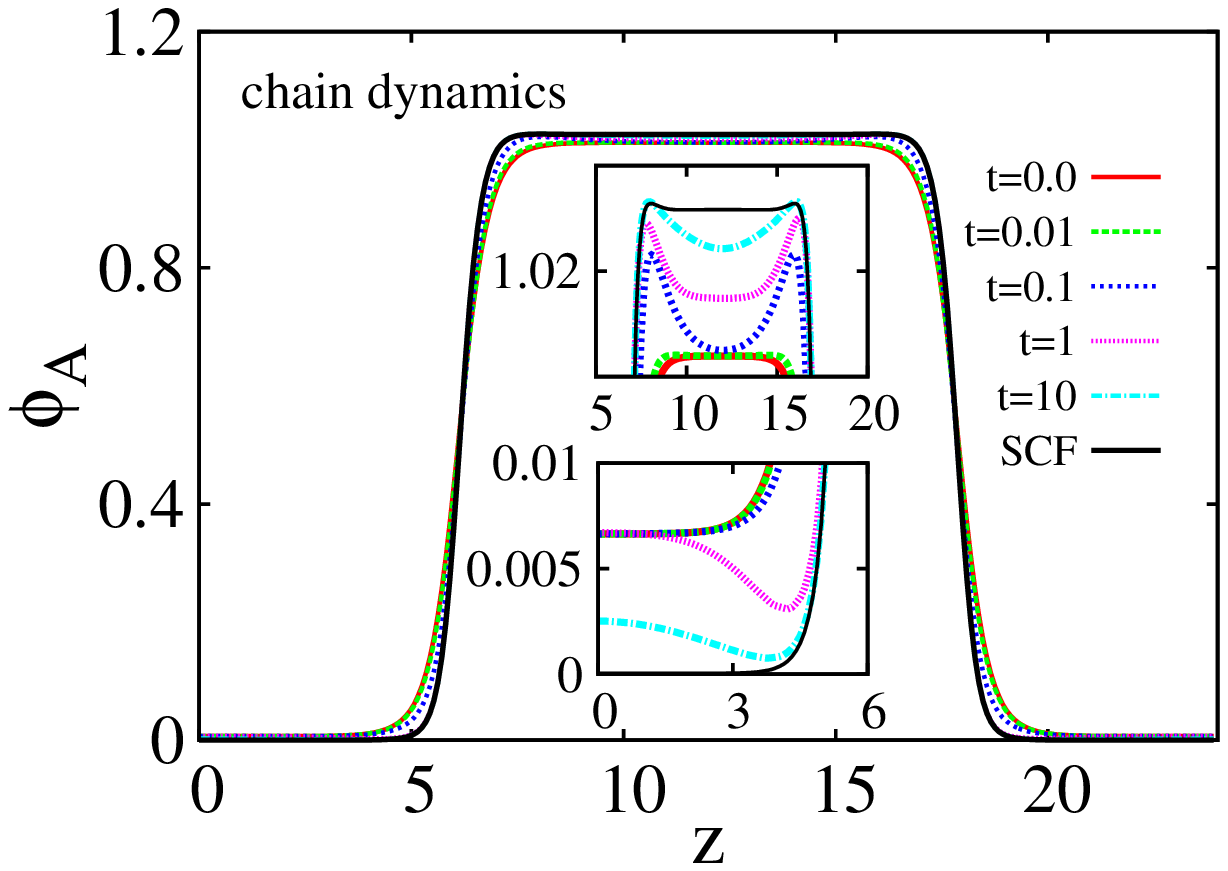}}
  \subfigure[]{
    \label{fig:6b} 
    \includegraphics[angle=0, width=7.0cm]{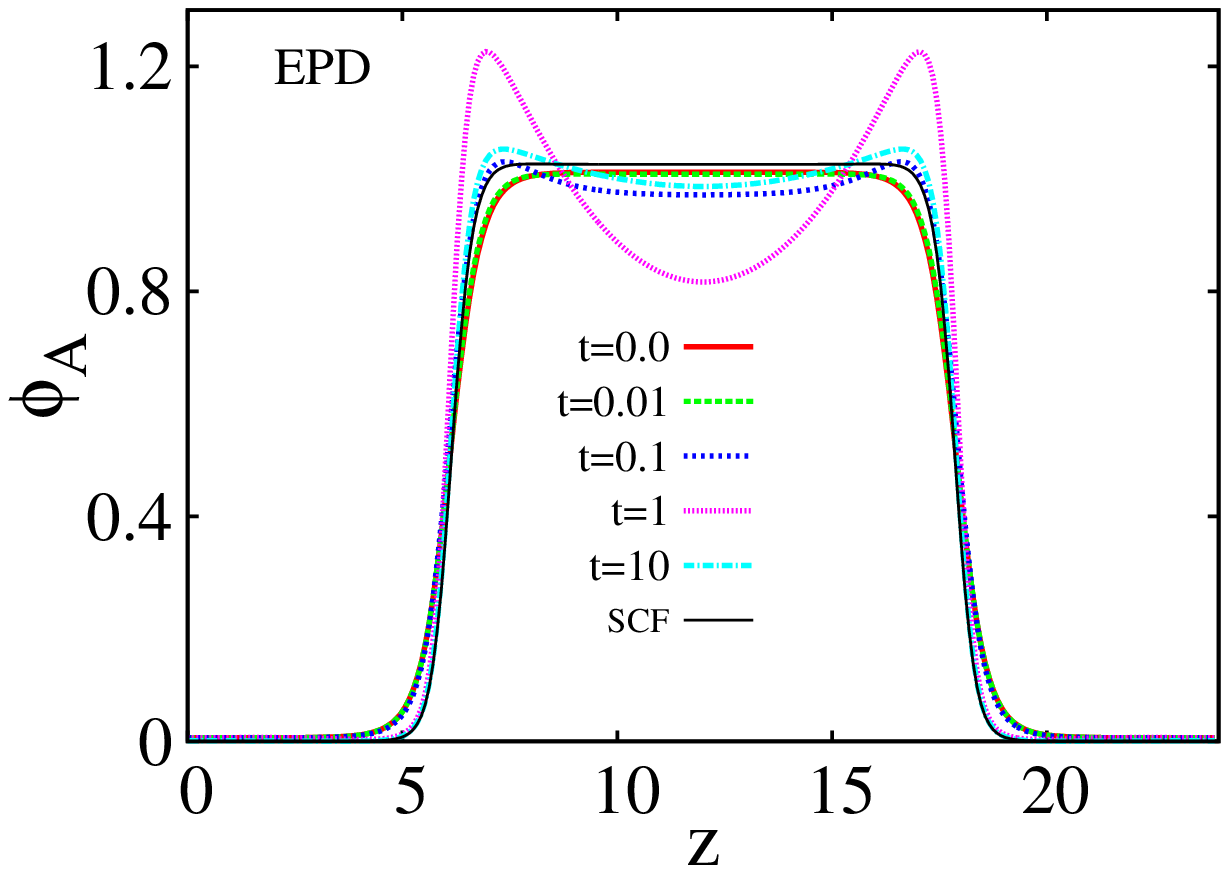}}
  \caption{Evolution of A-density profile during interfacial sharpening from
an interface corresponding to $\chi N = 5$ to an interface corresponding
to $\chi N = 10$, as obtained by chain dynamics DDF calculations (a)
and EPD calculations (b). The Two insets in (a) show blowups of some
parts of the curves. ``SCF" denotes the equilibrium density profile
according to self-consistent field theory at $\chi N = 10$.}
  \label{fig:density_sharpening} 
\end{figure}

Now we briefly consider the inverse problem, i.e. interface
sharpening after a sudden increase of $\chi N$. We choose as initial density
profile the equilibrium density distribution corresponding to $\chi N=5$, then
instantaneously quench the system to $\chi N=10$, and monitor the evolution of
the density profiles. As in the previous study (Figs.\ \ref{fig:density_tanh},
\ref{fig:width_tanh}), the system has the size $L_\mr z=24$ and is studied with
a spatial discretization corresponding to $n_\mr z = 256$ grid points. Fig.\
\ref{fig:density_sharpening}a) shows the density profiles at different times 
obtained from chain dynamics calculations. In the A-polymer-rich region,
the A-density gradually decreases to the final equilibrium value, and
in the A-polymer poor region, it gradually increases. Small transient peaks 
appear close to the interface, but no ``overshooting'' with respect to the
final equilibrium profile is observed. On the other hand, the EPD
calculations (Fig.\ \ref{fig:density_sharpening}b) predict large density
overshoots at times around $t \sim 1 t_0$ and a dramatic transient 
density reduction at the center of the A-rich slab. These phenomena
are obviously again artefacts of the EPD approximation.

We should note that the EPD method has been used to study interfacial
broadening in previous work, e.g., by one of us in Ref.\ \cite{interface_poly},
and no artefacts were observed. The reason is most likely that these studies
considered incompressible blends, where transient large density variations as
reported here were not possible. To further analyze the EPD artefacts, we
thus systematically study the effect of the Helfand compressibility parameter
$\kappa$ on the EPD prediction for the interfacial broadening problem.  The
parameters and initial conditions are chosen as in Fig.\ref{fig:width_tanh},
i.e., with a sharp initial interface. The Helfand parameter is increased from
$\kappa N=10$ to $\kappa N = 10.000$.  The time step in these simulations has
to be reduced with increasing $\kappa N$. Whereas $\Delta t = 10^{-3}t_0$ is
found to be sufficient at $\kappa N =10$ as explained above, we had to use
$\Delta t=10^{-6}t_0$ at $\kappa N = 10.000$.

\begin{figure}[htb]
  \centering
  \subfigure[]{
    \label{fig:7a} 
    \includegraphics[angle=0, width=7.0cm]{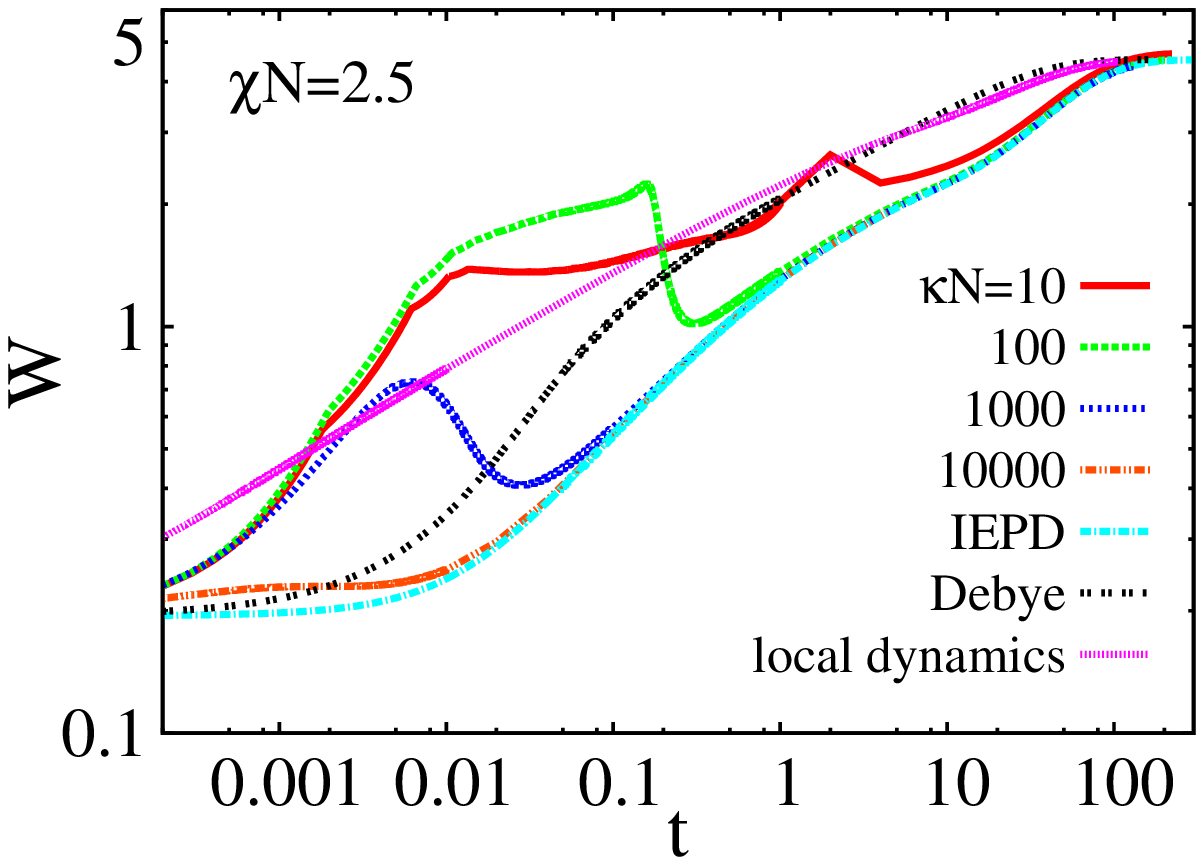}}
  \subfigure[]{
    \label{fig:7b} 
    \includegraphics[angle=0, width=7.0cm]{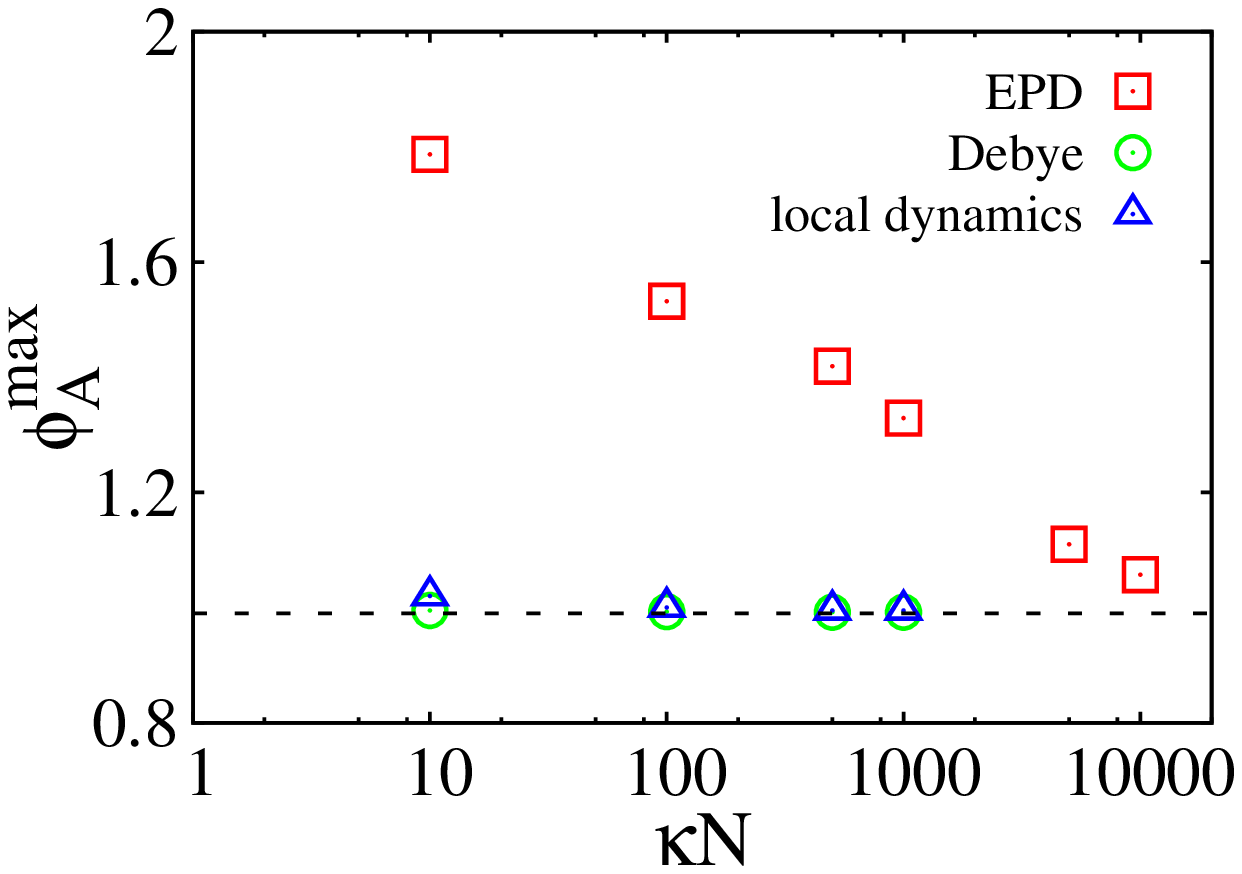}}
  \caption{(a) Interfacial broadening with respect to time according to
different DDF models and for EPD at different values of the Helfand
compressibility parameter $\kappa N$. IEPD refers to EPD in the incompressible
limit, $\kappa N \to \infty$. The results for Debye dynamics and local
dynamics for different $\kappa \in [10:1000]$ coincide; here we show the
curves for $\kappa N = 1000$. 
(b) Maximum density observed during the interfacial broadening
process for different DDF models as a function of compressibility parameter
$\kappa N$. The dashed line denotes the initial maximum density 
$\phi_A^\mr{initial} = 0.99$. 
The initial state is chosen as a sharp tanh function as in
Figure \protect\ref{fig:width_tanh} }.
 \label{fig:kappa} 
\end{figure}

Figure \ref{fig:7b} shows the maximum local density of A-monomers observed
during the whole course of a simulation, $\phi_A^\mr{max}$, as a function of
the compressibility parameter $\kappa N$. For EPD at $\kappa N = 10$, it is
almost twice as large as the mean {\em total} density $\bar{\phi} = 1$
(rescaled), which consistent with the density profile shown in Figure
\ref{fig:4b} (at $t = 1 t_0$). With increasing $\kappa N$, the maximum density
$\phi^\mr{max}_\mr A$ gradually decreases to reach $\bar{\phi}$ in the infinite
large $\kappa N$ limit.  As $\kappa N$ approaches infinity, the EPD dynamics
becomes similar to that of other DDFT models, i.e., the polymer density in the
polymer rich region decreases almost monotonically until it reaches the
equilibrium value.

Figure \ref{fig:7a} compares the EPD results for the time evolution of the
interfacial width for different Helfand parameters $\kappa N$. At $\kappa N
=10$, the curve exhibits several oscillations.  For larger $\kappa N$, it
becomes smoother, but remains nonmonotonic with a spurious peak at early times.
With increasing $\kappa N$, the peak becomes smaller and moves to earlier
times. For comparison, we also study the strictly incompressible limit $\kappa
N \to \infty$ as in Ref.\cite{interface_poly} (denoted IEPD). As expected, the
peak disappears. However, even in this limit, the EPD predictions do not
coincide with the results from Debye dynamics or local dynamics calculations
(the latter two are found to hardly depend on the compressibility parameter at
all). Considering the earlier finding that Debye dynamics calculations
reproduce the data from BD simulations nicely in Figure \ref{fig:width_chi} (at
$\kappa N =10$), we conclude that the EPD results should probably not be
trusted even in the incompressible limit, at least from a quantitative point of
view.

We have studied interfacial sharpening (Figure \ref{fig:6b}) for
larger $\kappa N$ (data not shown). Here again, the EPD artefacts disappear at
$\kappa N \to \infty$.  With increasing $\kappa N$, the spurious maxima in
$\phi_\mr A$ gradually decrease and vanish when $\kappa N$ approaches
infinity.


We close this subsection with a general remark on the Helfand parameter
$\kappa$.  It can be seen as a numerical convenience to replace a strictly
incompressible system. This is useful in continuum theories where the
incompressibility constraint makes the equations very stiff, and it is
necessary in particle-based simulations, where maintaining strict
incompressibility is close to impossible. In fact, strictly incompressible
systems do not exist in nature, even though typical values for the
compressibility parameter are of course much higher than $\kappa N = 10$.  In
our DDF calculations, we found the effect of compressibility on the interface
broadening dynamics to be negligible in all cases except EPD. Therefore, the
popular approach to approximate nearly incompressible blends by compressible
blends seems reasonable.

\subsection{Lamellar ordering in a diblock copolymer melt}
\label{sec:results_microphase}

\begin{figure}[htb]
  \centering
  \subfigure[]{
    \label{fig:8a} 
    \includegraphics[angle=0, width=7.0cm]{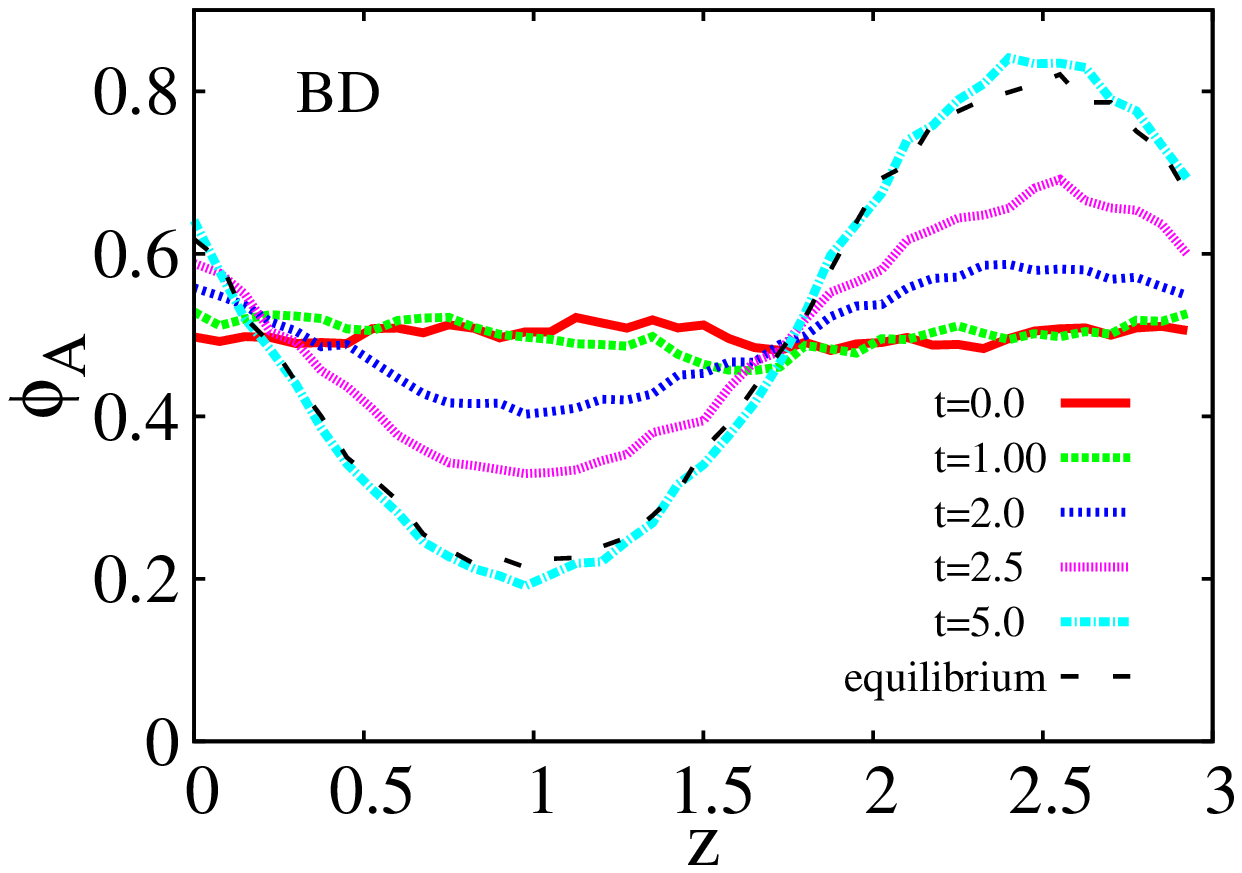}}
  \subfigure[]{
    \label{fig:8b} 
    \includegraphics[angle=0, width=7.0cm]{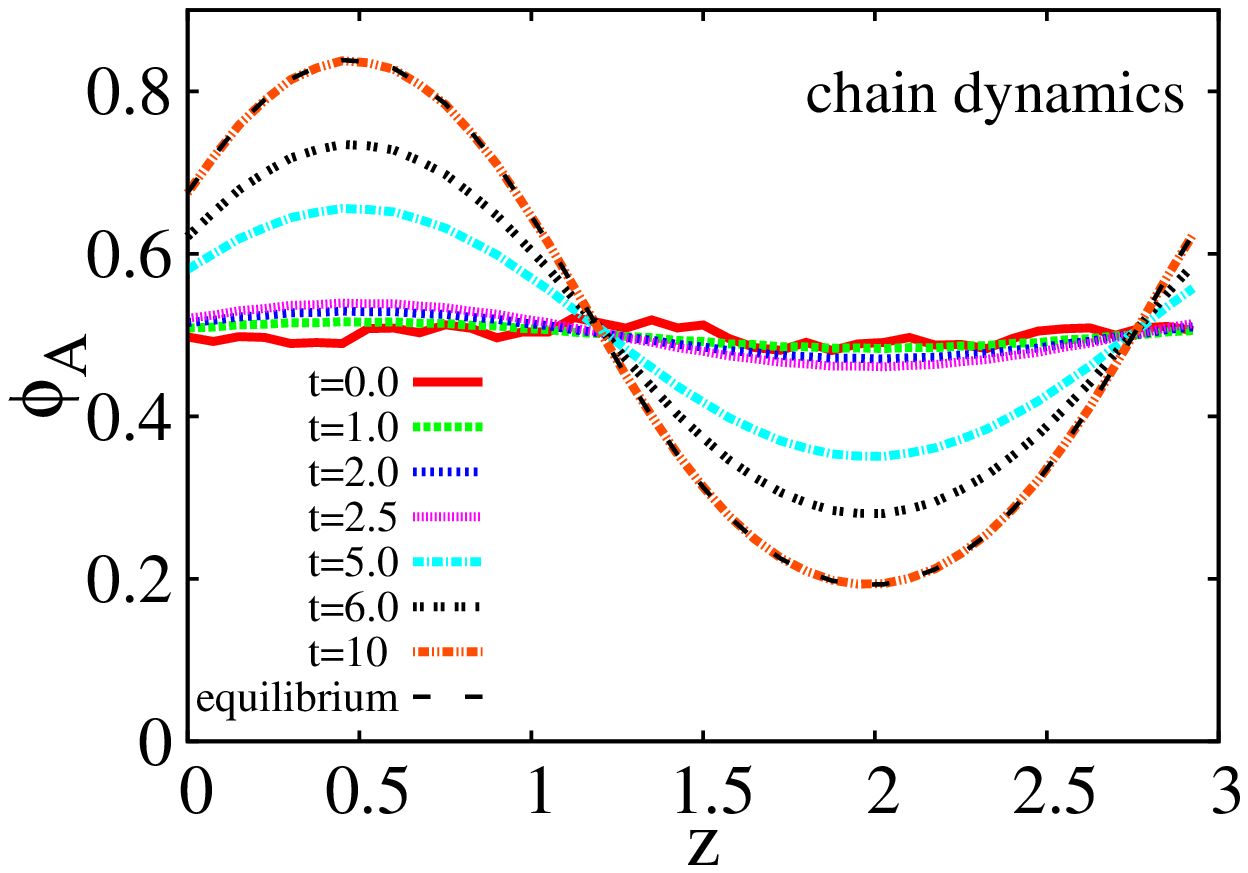}}
  \caption{Evolution of density profiles for the A monomers in A:B diblock
copolymer melts as obtained from BD simulations (a) and chain dynamics
calculations (b). The parameters are chosen $\chi N=12$, $\kappa N=10$, $N_\mr
A=N_\mr B$. They start with the same initial conditions. The time is measured
in units of $R^2_g/D_c$}
  \label{fig:evolution_copolymer} 
\end{figure}

Last, we consider the dynamics of microphase separation in initially disordered
copolymer melts. Experimentally, this has been studied, e.g., by Floudas
and coworkers \cite{lamellar_ordering_1, lamellar_ordering_2} and by Sakamoto
and Hashimoto \cite{lamellar_ordering_3} with small angle neutron and X-ray
scattering and other methods. These studies focussed on slow processes (time
scales of seconds) related to the nucleation and reordering of domains, and 
fluctuation effects in the vicinity of the order-disorder transition
\cite{lamellar_ordering_2}. Here we examine the kinetics of local
spontaneous ordering after a sudden deep quench into the ordered regime, on
time scales of submicroseconds that are hard to resolve experimentally. (For
typical polymers, the time scale $t_0$ is roughly of the order $10^{-5}$
seconds.) 

We consider systems of $n_\mr c$ identical A:B diblock copolymers made of $N$
monomers, where the A block and B block have the same length, i.e. $N_\mr
A=N_\mr B$. In all calculations, we set $\chi N=12$, $\kappa N=10$, which lead
to a lamellar morphology at equilibrium. We initialize the system by imposing
weakly inhomogeneous density distributions, and monitor the evolution of the
A-density profile until it is equilibrated. In particular, we consider the
evolution of the A-density at the location where it assumes its maximum value
(which we denote $\phi_A^\mathrm{max}$) at the end of the run. The BD
simulations are implemented in three dimensional space in systems of size
$L_\mr x=L_\mr y=1$, and $L_\mr z=3$ with $n_\mr x\cdot n_\mr y\cdot n_\mr
z=10\cdot 10\cdot 40$ uniform cubic cells.  The resulting density profiles are
homogeneous along the $x$ and $y$ directions and exhibit a lamellar structure
with one period in the $z$ direction. We should note that the system is
slightly frustrated at equilibrium, as the bulk lamellar spacing is roughly 3.4
$R_g$, which is not fully commensurable with the box length $L_\mr z$. Since we
are using the same box dimensions in the BD simulations and the DDF
calculations, quantitative comparisons are still possible.

Specifically, we studied systems of 5000 chains of length $N=40$. For
comparison, we have also considered systems with chain length $N=20$ and $N=60$
at the same mean monomer density (i.e., 10.000 chains and 3000 chains), and
obtained similar results (data not shown). The average number of monomers per
cell is about 50. The time step in the BD simulations is chosen $\Delta
t=0.0003/N$ (i.e., $\Delta t = 0.75 \cdot 10^{-5} t_0$ at $N=40$).  The DDF
calculations are carried out in one dimension with time step $\Delta t = 10^{-4}
t_0$ in the EPD and Debye dynamics models, $\Delta t = 10^{-5} t_0$ in
the chain dynamics model, and $\Delta t = 10^{-6} t_0$ in the local
dynamics model. We found that such small time steps were necessary in
the chain dynamics and local dynamics schemes, otherwise the iterative
reconstruction of the auxiliary fields $\omega_\alpha$ from the density fields
$\phi_\alpha$ did not always converge. Adaptive time steps would probably
partly relieve these problems.

Fig.\ \ref{fig:evolution_copolymer} shows the density profiles of A-monomers at
different times. In the BD simulations, the initial polymer configurations are
generated as free Gaussian chains which creates a random density distribution.
This distribution, averaged over the $x$ and $y$ directions, is imposed as
initial density profile in all DDF schemes. Indeed, we can see from
Fig.\ref{fig:evolution_copolymer}a and b that $\phi_A(z)$ are the same at
$t=0$. As time passes, the system microphase separates into A-rich and B-rich
regions and reaches equilibrium at large $t$. The shapes of the equilibrium
densities can almost be superimposed after performing a proper translational
shift along the abscissa, they therefore represent the same equilibrium state.
However, the phase ordering proceeds at different speeds.  For example,
focussing on the maximum density $\phi_\mr A^\mr{max}$ at $t=2$, we can see
that the BD simulations give $\phi_\mr A^\mr{max}\simeq 0.59$ at this time
(Fig.\ \ref{fig:evolution_copolymer}a), while the chain dynamics calculation
predicts $\phi_\mr A^\mr{max}\simeq 0.51$ (Fig.\
\ref{fig:evolution_copolymer}b).  Hence the ordering in the chain dynamics
model is too slow. The curves obtained from other DDF models
(including EPD) are qualitatively similar to the BD curves, but the density
evolves at different speeds (data not shown).

\begin{figure}[htb]
  \centering
  \subfigure[]{
    \label{fig:9a} 
    \includegraphics[angle=0, width=7.0cm]{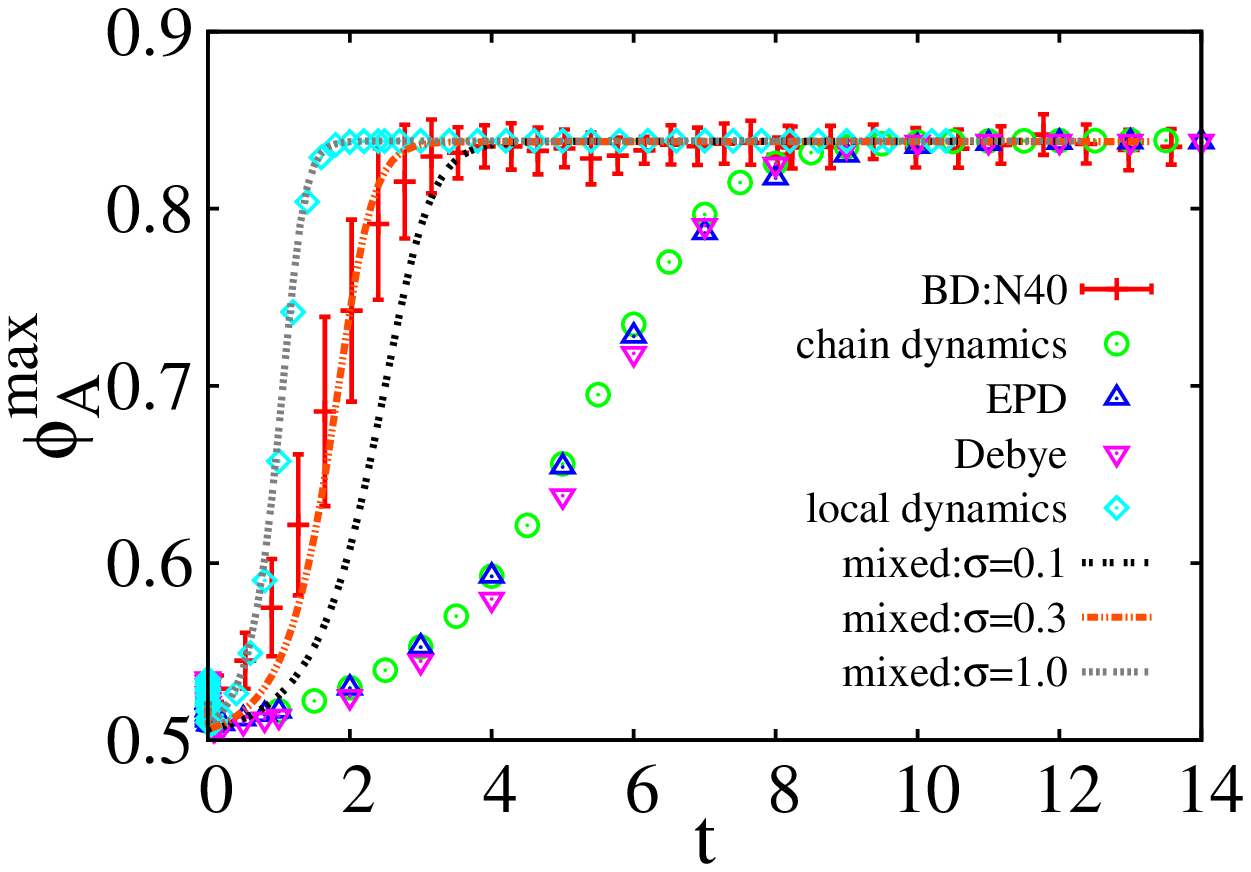}}
  \subfigure[]{
    \label{fig:9b} 
    \includegraphics[angle=0, width=7.0cm]{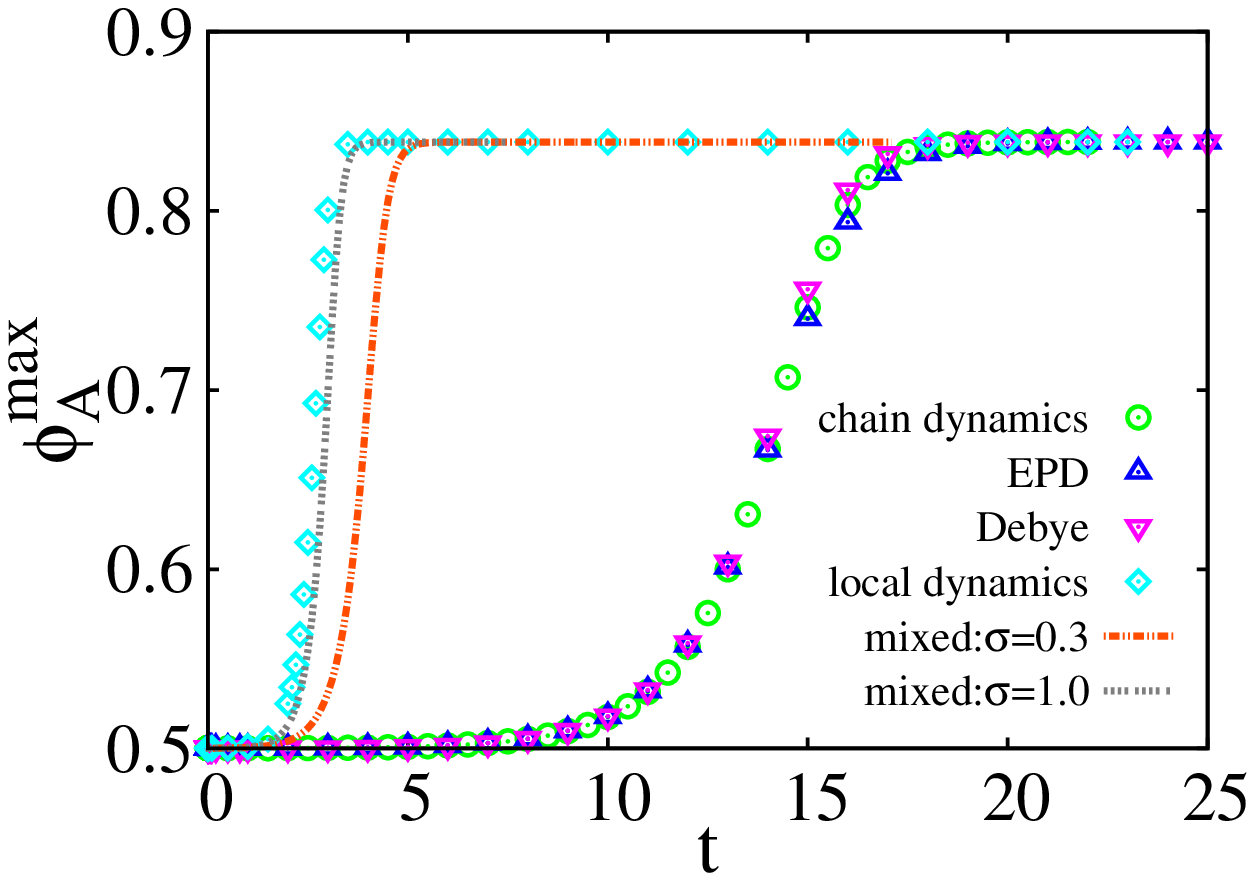}}
  \caption{Evolution of $\phi_A^\mathrm{max}$ with time as obtained from
BD simulations and the field-based DDF schemes discussed in this work. 
In (a), the DDF profiles are initialized to precisely match the initial 
density distribution in the BD simulations for $N=40$,
while in (b), they are initialized by imposing a very small random
noise to the initial auxiliary potentials $\omega_\ga(\rr)$. (see text
for explanation}
  \label{fig:density_max} 
\end{figure}

To further quantify this observation, we now focus on $\phi_A^\mathrm{max}$.
Fig.~(\ref{fig:density_max}a) compares the evolution of $\phi_A^\mathrm{max}$
with respect to time obtained from BD simulations with the results of the
different DDF models, choosing as initial DDF condition the BD monomer
distribution profile at $t=0$. For such ``realistic'' initial condition,
$\phi^{\mathrm{max}}_\mr A \simeq 0.53$ at time $t = 0$, indicating that the
inhomogeneities in the initial (fully disordered) system are already very high.
As time progresses, $\phi^\mathrm{max}_\mr A$ first decreases slightly, and
then increases continuously until it saturates at the equilibrium value. The
predictions from nonlocal chain dynamics models (chain dynamics, Debye
dynamics, and EPD) are in good agreement with each other, but they
underestimate the rate of ordering. In contrast, the local dynamics model
overestimates the ordering velocity. At early times ($t \le 1$), the local
dynamics prediction almost matches the results from the reference BD
simulations.  Thus the ordering of monomers seems to dominate the dynamics on
time scales smaller than the characteristic relaxation time of a chain. At
later times, however, the diffusion of the whole chain becomes significant, and
large scale correlations begin to play a role.  It is then necessary to account
for the nonlocal correlations that slow down the dynamics. The density
evolution predicted by the pure local dynamics scheme therefore overestimates
the rate of ordering at late times. 

This problem is addressed by the mixed DDF scheme introduced in Sec.\
\ref{sec:DDF}, Eq.\ (\ref{eq:ddf_mixed}). In this scheme, a tunable parameter
$\sigma$ is used to control the crossover between local and nonolocal dynamics.
On small length scales below $\sigma$, the dynamics is mostly local, whereas on
larger scales, it becomes nonlocal. The effect can be seen in Fig.\
\ref{fig:density_max}a). For $\sigma=1$, the curves calculated with the mixed
dynamics reproduce the local dynamics curves. For $\sigma=0.1$, the evolution
of $\phi^\mathrm{max}_A$ is slow and stays on the side of nonlocal dynamics.
However, at $\sigma=0.3$, the mixed dynamics approximately reproduces the
BD simulation data: If one propagates the motion according to local dynamics on
length scales smaller than $0.3 R_g$, and according to nonlocal dynamics on
length scales larger than $0.3 R_g$, one obtains a dynamical scheme that is
quite close to describing the true dynamics.  This mixed dynamics scheme
accounts for the fact that in the phase ordering process, both the small-scale
motion and the large-scale motion are important. The parameter $\sigma$ roughly
specifies the weight of these two types of dynamics. 

The results of the DDF calculations depend sensitively on the choice of the
initial conditions. This is demonstrated in Fig.\ \ref{fig:density_max}b),
which shows that the DDF results if one starts with an almost homogeneous
initial density profile decorated with a very weak unphysical noise does not
have the correct spatial correlations of a disordered copolymer melt.
Specifically, the initial conditions are defined through the initial auxiliary
potential fields, which are chosen $\omega_\ab(\rr) = 0.01 \: \xi(\rr)$, where
$\xi$ are uniformly distributed random numbers in the interval $\xi \in [0:1]$.
This results in an initial maximum A-density of $\phi^\mathrm{max}_\mr A
\approx 0.5002$. The subsequent time evolution of the density profiles is
qualitatively similar to that obtained with realistic initial conditions (Fig.\
\ref{fig:density_max}a): The ordering is much slower in the nonlocal DDF
schemes than in the local DDF scheme. The large difference again reflects the
multiscale character of the phase ordering process. Most importantly, the
comparison with Fig.\ \ref{fig:density_max}a) shows that the onset of ordering
can be delayed significantly if the initial conditions are not chosen
appropriately. The problem can presumably be reduced by including thermal noise
as described at the end of Sec.\ \ref{sec:DDF}. This has not been done here.

We have also examined the effect of compressibility on the ordering
kinetics. We found that increasing $\kappa N$ affects the final density
distributions slightly, but it has a negligible effect on the ordering kinetics
(data not shown). For example, when examining the maximum density of A
monomers, $\phi^\mr{max}_\mr A$, as a function of time $t$, the curves for
different $\kappa N$ (using the same DDF theory) overlap almost completely
during the ordering process, but they saturate to different equilibrium
values.

\section{Conclusion and Outlook} \label{sec:summary}

The main results of the present work can be summarized as follows: We have
compared the predictions of different DDF theories with BD simulations of
interface broadening in compressible homopolymer blends and microphase
separation in copolymer melts.  

\begin{itemize}

\item Interface broadening in blends is best described by the nonlocal
``chain dynamics'' DDF theory. 

\item DDF calculations based on the Debye approximation do not differ
significantly from the full chain dynamics calculations in all situations
considered here.  In contrast, when looking at interface broadening or
sharpening, EPD calculations produce spurious artefacts at intermediate times.
The problem becomes worse if the interfaces are sharper. The artefacts
disappear in the incompressible limit, however, the EPD results still differ
noticeably from those of other DDF theories. This must be attributed to the
EPD approximation and is most likely a consequence of the fact that the EPD
model does not strictly guarantee local mass conservation.

\item Neither local dynamics nor chain dynamics can capture the kinetics of
microphase separation in block copolymer melts. Compared to the reference BD
simulations, local dynamics calculations underestimate the ordering time, and
chain dynamics calculations overestimate it. This most likely reflects the
multiscale character of the ordering process, which involves both local chain
rearrangements and global chain motions.  To address this problem, we have
proposed a mixed local/nonlocal DDF scheme, which combines local monomer motion
on small scales below $R_g$ with global cooperative diffusion on large length
scales. This scheme can reproduce the BD simulation results at an almost
quantitative level for all situations considered in the present work.

\end{itemize}

Our mixed DDF approach has some similarity to DDF approaches that have been
proposed in the 90s for studying reptation dynamics in strongly entangled
polymer systems \cite{Kawasaki2,Harden}. Here, an effective nonlocal Onsager
matrix is also constructed such that monomers move differently from whole
chains. In the reptation models, the mobility of single monomers is assumed to
be reduced compared to whole chains due to their confinement to a tube. Our
results here indicate, for the Rouse regime, that it is rather enhanced. 

All these DDF schemes have been postulated more or less heuristically.
However, our mixed scheme contains one free parameter (the ``filter''
parameter $\sigma$), which can be used to adjust the DDF calculations to the BD simulations.
On the one hand, this reduces the ``predictive power" of the approach. On the other hand, it offers
a way to incorporate information from more detailed fine-grained models for polymer dynamics in a DDF model in
a coarse-grained sense. The parameter $\sigma$ can then be seen as an effective parameter in a dynamic
field theory for polymers, which might have to be determined from fine-grained simulations and experiments. In future work,
we thus plan to systematically construct mixed schemes from
fine-grained simulations, e.g., based on dynamic correlation functions in
reference particle simulations. We hope that this approach will help to obtain
a more accurate description of polymer dynamics at the field-based level,
without the need of explicitly accounting for the multiple time scales involved
in polymer relaxation and the corresponding memory effects. Ideally, it should
not be restricted to polymers in the Rouse regime, but could also be applied to
entangled polymer systems or other complex fluids, and possibly even to
nonequilibrium systems under shear stress where polymers are deformed
\cite{Mueller_Tang}.

To summarize, in the present work, we have evaluated different dynamic
density functional (DDF) theories for the description of kinetic processes in
inhomogeneous polymer systems. As mentioned in the introduction, {\em static}
density functional theories have proven to be very successful and powerful tools
for predicting self-assembled polymeric nanostructures at equilibrium.
However, in practice, self-assembly is a nonequilibrium process which often
does not run to completion. Polymeric nanostructures are usually not fully
equilibrated, and their morphologies and even characteristic length scales may
strongly depend on the history of the self-assembly \cite{selfassembly,Simon}.
This is in fact an advantage, because process design can be used as an
additional design principle.  However, it also implies that the resulting
structures cannot be predicted by static density functional theory alone.  We
believe that systematic assessments of DDFs such as the one presented here are
necessary steps towards an improved theoretical description of nonequilibrium
{\em dynamic} processes and the resulting nonequilibrium structures.

\bigskip
\begin{center}
\textbf{ACKNOWLEDGMENTS}
\end{center}

Financial support from the German Science Foundation (DFG) within project C1 in
SFB TRR 146 is gratefully acknowledged. Simulations have been carried out on
the computer cluster Mogon at JGU Mainz.

\begin{appendix}

\section{Force calculation in BD simulations}\label{appendix:BD}

In this appendix, we give the explicit expression of the potential force acting
on each bead, using the polymer A/B blend as the model system. The simulation
box is uniformly divided into $n_\mr x\cdot n_\mr y\cdot n_\mr z$ cells, and
densities are defined on the vertexes (mesh points) of these cells. Each cell
has a volume of $l_x\cdot l_y\cdot l_z$ (with $l_I=L_I/n_I$). Fractions of a
bead are assigned to its neighbouring mesh points according to predefined
assignment functions $h(r)$ that depend only on the distance between the
particle and the mesh point. Any small displacement of a bead causes a density
change, and hence a change of the Hamiltonian. Therefore, the bead experiences
a force.  In the following, we focus on the (non-bonded) interaction part of
the Hamiltonian, and rewrite this part in a discretized form as
\begin{eqnarray}
H_\mr I&=&\frac{n\chi N}{V}\sum_g\Delta V\hat\phi_A(\mb r_g)\hat\phi_B(\mb r_g)\nonumber\\
&+&\frac{n\kappa N}{2V}\sum_g\Delta V\big[\hat\phi_A(\mb r_g)+\hat\phi_B(\mb r_g)-1\big]^2,
\end{eqnarray}
where $g$ denotes the index of the mesh point on which the densities are
defined, and $\Delta V=l_xl_yl_z$ is the volume of a cell. The densities are
calculated using an assignment function, i.e., $\hat\phi_\alpha(\mb
r_g)=\frac{1}{\Delta V\rho_0}\sum_jh(|\mb R_j-\mb r_g|)$ where $\mb R_j$ is the
position of the j-th bead, and j runs over all beads of type $\alpha$. We
consider the force acting on an A bead at position $\mb R=(x,y,z)$. The
derivative of $H^{(1)}$ with respect to $\mb R$ can be written as
\begin{eqnarray}
\frac{\partial H_\mr I}{\partial\mb R} &= &
   \frac{\partial h(|\mb R-\mb r_g|)}{\partial\mb R}
   \Big[
   \chi
   \sum_g\hat\phi_B(\mb r_g)
   \nonumber \\
  && + 
  \kappa
    \sum_g\Big(\hat\phi_A(\mb r_g).
    +\hat\phi_B(\mb r_g)-1\Big) \Big]
\end{eqnarray}

In order to proceed, we need to give an explicit expression for the assignment
function. For this purpose, we consider the cell where $\mb R$ is located. The
cell has eight vortices at the corners, which we label by indices  $i,j,k$
along $x,y,z$ directions, such that the set of indices $i=0, j=0, k=0$ marks
the vertex number 0 with coordinate $(0,0,0)$, $i=0, j=0, k=1$ marks the vertex
number 1 with coordinate $(0,0,l_z)$, $i=0, j=1, k=0$ marks the vertex number 2
with coordinate $(0,l_y,0)$, and so on until $i=1,j=1,k=1$ marks the the vertex
number 7 with coordinate $(l_x,l_y,l_z)$. Several choices for the assignment
function are conceivable. In the lowest order scheme -- the so-called
nearest-grid scheme -- each bead is fully assigned to its nearest mesh point. Here we
use a higher order scheme, which assigns fractions of each bead to its eight
nearest mesh points. The fraction assigned to a given vertex is proportional to
the volume of a rectangle whose diagonal is the line connecting the particle
position and the mesh point on the opposite side of the mesh cell. With the
precise arrangement of vortices, the assignment function for each vertex can be
written as
\begin{equation}
h(|\mb R-\mb r_g|)
  =\frac{(l_x-|r_{gx}-x|)(l_y-|r_{gy}-y|)(l_z-|r_{gz}-z|)}{l_xl_yl_z},
\end{equation}
where $g$ ranges from 0 to 7, and $r_{g\alpha}$ is the $\alpha$ component of
$\mb r_g$. With these assignment functions, the force from non-bonded
interactions acting on an A bead along the $x$ direction is given by
\begin{eqnarray}
-F_x&=&\frac{\partial H_\mr I}{\partial R_x}
  =\frac{1}{N}\frac{\partial u_A}{\partial R_x}
  =\frac{ u_A(\mb r_5)-u_A(\mb r_1)}{Nl_x}\frac{(l_y-y)z}{l_yl_z}
  \nonumber\\
  &+& \frac{u_A(\mb r_6)-u_A(\mb r_2)}{Nl_x}\frac{y(l_z-z)}{l_yl_z}
   +\frac{u_A(\mb r_7)-u_A(\mb r_3)}{Nl_x}\frac{yz}{l_yl_z}
  \nonumber\\
  &+&\frac{u_A(\mb r_4)-u_A(\mb r_0)}{Nl_x}\frac{(l_y-y)(l_z-z)}{l_yl_z},
\end{eqnarray}
the non-bonded force in $y$ direction is
\begin{eqnarray}
-F_y&=&\frac{\partial H_\mr I}{\partial R_y}=\frac{1}{N}\frac{\partial u_A}{\partial R_y}
  =\frac{u_A(\mb r_6)-u_A(\mb r_4)}{Nl_y}\frac{x(l_z-z)}{l_zl_x}
  \nonumber\\
  &+&\frac{u_A(\mb r_3)-u_A(\mb r_1)}{Nl_y}\frac{(l_x-x)z}{l_xl_z}
    +\frac{u_A(\mb r_7)-u_A(\mb r_5)}{Nl_y}\frac{zx}{l_xl_z}
  \nonumber\\
&+&\frac{u_A(\mb r_2)-u_A(\mb r_0)}{Nl_y}\frac{(l_x-x)(l_z-z)}{l_xl_z}
\end{eqnarray}
and the non-bonded force in $z$ direction is
\begin{eqnarray}
-F_z&=&\frac{\partial H_\mr I}{\partial R_z}=\frac{1}{N}\frac{\partial u_A}{\partial R_z}
   =\frac{u_A(\mb r_5)-u_A(\mb r_4)}{Nl_z}\frac{x(l_y-y)}{l_xl_y}
 \nonumber\\
  &+&\frac{u_A(\mb r_3)-u_A(\mb r_2)}{Nl_z}\frac{(l_x-x)y}{l_xl_y}
   +\frac{u_A(\mb r_7)-u_A(\mb r_6)}{Nl_z}\frac{xy}{l_xl_y}
  \nonumber\\
  &+&\frac{ u_A(\mb r_1)-u_A(\mb r_0)}{Nl_z}\frac{(l_x-x)(l_y-y)}{l_xl_y}.
\end{eqnarray}
Here $R_\mr I$ denotes the I-component of $\mb R$, while $F_\mr I$ is the Ith
component of the force $\mb F$. Similar expressions are obtained for the
non-bonded interaction forces acting on B beads.

\section{Discussion and Integration of nonlocal DDF equations}
\label{appendix:field_dynamics}

In the following, we briefly sketch the derivation of the DDF models introduced
in Sec.\ \ref{sec:DDF} and present the numerical method which we use to
integrate the chain dynamics equation.  We derive the DDF equations of the
non-local chain dynamics DDF model following Maurits \etal \cite{EPD1997},
at the example of A:B diblock copolymer melts. The extension to other polymer
systems is straightforward.  

Our system contains $n_\mr c$ diblock copolymers in a volume $V$. Each chain
has the length $N=N_\mr A+N_\mr B$, where $N_\mr A$ is the length of block $A$
and $N_\mr B$ the length of block $B$. For convenience, we choose the
continuous chain model, where the chain length is scaled to 1, i.e., $1=N_\mr
A/N+N_\mr B/N\equiv h_\mr A+h_\mr B$, and we define the sequence function
$\tau(s)=A$ for $s < h_A$ and $\tau(s)=B$ for $h_A<s<1$. The units of length,
energy, and time are chosen as in the main text, i.e., we set $k_BT\equiv1$ as
the energy unit, and the radius of gyration of a free ideal chain $R_\mr
g=a\sqrt{N/6}$ as the length unit ($a$ is the statistical Kuhn length), and
measure time in units of the relaxation time of the whole chain,
$\tau=R_g^2/D_\mr c$, where $D_\mr c$ is the diffusion constant of a whole
chain. 

In the Rouse regime, the internal structure of chains relaxes faster than the
coarse-grained collective motion, therefore whole chains are taken to drift
with uniform velocity, according to the following equation of motion:
\begin{equation}
\frac{\ud \mb R(s,t)}{\ud t}
   = N \int_0^1 \ud s \sum_{\gb=A,B} f_\gb[\mb R(s,t)] 
      \: \delta_{\beta, \tau(s)}.
\end{equation}
Here $f_\gb=-\nabla \tilde{\mu}_\gb$ is the thermodynamic force acting on a
monomer of type $\gb$ (which is determined from the Helmholtz free energy
functional (\ref{eq:scf_free_energy}) via  $\tilde{\mu}_\gb=\delta F/\delta
\phi_\gb$), $s$ is the contour variable, and  $\mb R(s,t)$ denotes the
conformation of the chain at time $t$. The conformation dependent rescaled
density is given by $\hat\phi_\ga=\rho^{-1}_0 N\sum_m\int_0^1 \ud s\delta(\mb
r-\mb R_m(s,t)) \delta_{\ga,\tau(s)}$, where the index $m$ runs over all
$n_{\mr c}$ copolymers, and $\rho_0=n_{\mr c}N/V$ is the reference density. The
evolution equation for the rescaled density can be obtained by taking directly
its derivative with respect to time. After taking the ensemble average at both
sides, we obtain the dynamical equation
\begin{equation}
\label{eq:ddf2}
\partial_t \phi_\ga
  = \nabla\cdot\int d\mb r'\sum_\gb\Big[\tLCD_\ab(\rr,\rr',t) 
       \nabla'\tilde{\mu}_\gb(\rr')\Big]
\end{equation}
with $\phi_\alpha = \langle\hat\phi_\alpha\rangle$, where
$\langle\cdots\rangle$ refers to the ensemble average. This is exactly 
Eq.\ (\ref{eq:ddf_general}), with Onsager coefficients (the
correlators) defined as 
\begin{eqnarray}
\lefteqn{\tLCD_\ab(\rr,\rr',t)\equiv\rho_0^{-1} n_\mr c N} && \nonumber\\
  &&\times \: \int_0^1 \ud s \int_0^1 \ud s'  
  \Big\langle \delta(\rr-\mb R(s,t)) \delta(\rr'-\mb R_\gb(s',t)) 
       \Big\rangle \nonumber\\
  &=&-\frac{\delta\phi_\ga(\rr,t)}{\delta\omega_\gb(\rr',t)}.
\label{eq:ddfchain}
\end{eqnarray}

Eq.\ (\ref{eq:ddf2}) is used to derive a set of approximate dynamic schemes
including the EPD scheme and the Debye scheme.  The approximation involved in
EPD is translational symmetry, i.e., one assumes $\nabla \tLL_\ab(\rr,\rr')
\simeq-\nabla'\tLL_\ab(\rr,\rr')$. Employing this assumption and using the
chain rule, $\partial_t \phi_\ga=\int \ud \rr' \sum_\gb
\frac{\delta\phi_\ga}{\delta\omega_\gb(\rr')} \partial_t \omega_\gb(\rr')$, one
can transform the density evolution equation into an equation propagating the
``potential fields''
\begin{equation}
\partial_t \omega_\ga =-\nabla^2\mu_\ga.
\end{equation}
Thus the dynamic equations are simplified considerably. They can be integrated
conveniently in Fourier space using fast Fourier transform (FFT). One big
advantage of the EPD scheme is that the computationally cumbersome inverse
determinations of potential fields $\{\omega_\ga\}$ from density fields
$\{\phi_\ga\}$ are avoided, since the $\omega_\ga$ are propagated directly.

The Debye scheme is obtained by applying a weak inhomogeneity expansion (random
phase approximation, RPA) \cite{polymer_book}, where the true correlations are
replaced by the correlation functions of ideal Gaussian chains, i.e.,

\begin{equation}
\label{eq:debye1}
\tLDB_\ab(\rr,\rr',t)\simeq g_\ab^D(\rr-\rr'),
\end{equation}
where $g_\ab^D$ is best given in Fourier space $\qq$ with
\begin{eqnarray}
\label{eq:debye2}
g_{\ga \ga}^D(\qq) & = & g_D(h_\ga,x) \\
\label{eq:debye3}
g_{AB}^D(\qq) & = & \frac{1}{2} 
  \Big( g_D(1,x)- g_D(h_A,x) - g_D(h_B,x) \Big),
\end{eqnarray}
with $x=q^2 R_g^2$ and the Debye function $g_D(h,x) = \frac{2}{x} (hx+{\rm
e}^{-hx}-1)$.  In the case of A/B  homopolymer blends made of $n_\ga$
homopolymers $\ga$, Eqs.\ (\ref{eq:debye1}-\ref{eq:debye3}) are replaced by 
\begin{equation}
\label{eq:debye4}
\tLDB_{\ga \ga} = \frac{n_\ga}{n_\mr A+n_\mr B} \: g_D(1,q^2 R_g^2), \quad
\tLDB_{AB} \equiv 0. 
\end{equation}
In the Debye approximation, the DDF equation for the $\alpha$-component 
can be written in Fourier space as
\begin{equation}
\partial_t \phi_\ga = -q^2\sum_\gb \tLDB_\ab (\qq) \: \tilde{\mu}_\gb(\qq),
\end{equation}

Finally in this appendix, we will now present our numerical method for 
propagating the densities according to the full chain dynamics DDF equations,
(\ref{eq:ddf2}) with (\ref{eq:ddfchain}), without further approximations. We
first define the auxiliary vector valued field 
\begin{equation}\label{eq:intermediate}
 {\mb V}_\ga (\rr)=-\int \ud \rr' \sum_\gb 
      \tLCD_\ab(\rr,\rr') \nabla'_\gb \tilde{\mu}_\gb(\rr').
\end{equation}
Using this intermediate variable, we can rewrite the DDF equation as
\begin{equation}
\label{eq:intermediate2}
\partial_t \phi_\ga(\rr,t) = - \nabla {\mb V}_\ga(\rr,t).
\end{equation}
Provided that ${\mb V}_\ga$ is known, the integration of this equation is
straightforward. Thus we are left with the task to evaluate the intermediate
field ${\mb V}_\ga(\rr,t)$. To this end, we make use of the relation
$\tLCD_\ab(\rr,\rr') = - \delta \phi_\ga(\rr)/\delta \omega_\gb(\rr')$ and
rewrite Eq.\ (\ref{eq:intermediate}) as
\begin{eqnarray}
 {\mb V}_\ga (\rr)&=&\int \ud \rr' \sum_\gb 
  \delta \phi_\ga(\rr)/\delta \omega_\gb(\rr')
        \nabla'_\gb \tilde{\mu}_\gb(\rr')
 \nonumber\\
 & = & \nabla_{\mb u}
  \phi_\ga [\{ \omega_\gb + {\mb u} \cdot \nabla \tilde{\mu}_\gb\} ]_
    {{\mb u}=0},
\label{eq:intermediate3}
\end{eqnarray}
where $\nabla_{\mb u}$ denotes the gradient operator with respect to ${\mb u}$.
Eqs.\ (\ref{eq:intermediate2}) and (\ref{eq:intermediate3}) suggest the
following Euler forward algorithm for integrating the chain dynamics DDF 
equations:

\begin{enumerate}

\item Find the initial potential $\omega_\alpha^{(0)}$ corresponding 
to the initial density $\phi_\alpha^{(0)}$.
This is done by numerical iteration methods.  

\item Choose a small parameter $\epsilon$, and calculate 
the components $I=x,y,z$ of ${\mb V}_\ga$ according to

\begin{equation}
V_{\ga,I} = \frac{1}{\epsilon}
  \phi_\ga[\{\omega_\gb^{(0)}+\epsilon \partial_I \tilde{\mu}^{(0)}_\gb\}]
    -\phi_\ga[\{\omega_\gb^{(0)}\}],
\end{equation}

where $\tilde{\mu}^{(0)}_\ga$ is an explicit function of
$\{\omega_\gb^{(0)}\}$ and $\{\phi_\gb^{(0)}\}$. 

\item Propagate the density over one time step, i.e., evaluate the density at
time $t+\Delta t$ using the explicit Euler scheme

\begin{equation}
 \phi_\ga^{(1)}(t+\Delta t)
 =\phi_\ga^{(0)}(t)-\Delta t \: \nabla \cdot {\mb V}_\ga.
\end{equation}

\end{enumerate}

The above procedure is repeated to obtain the time evolution of the densities
as well as the auxiliary potentials. The same idea can be used to construct
more sophisticated integration schemes (beyond explicit Euler).  Compared to
the EPD scheme, the present scheme is more accurate, since it does not rely on
the EPD assumption of translational symmetry. However, it requires much more
computing time, since it involves the evaluation of all spatial components of
${\mb V}$ and the iterative reconstruction of the auxiliary potentials from the
densities in each time step.

\section{Dynamic density functional theory with fluctuations}\label{appendix:fluctuations}

On the basis of DDFT, fluctuations can be included
in Eq.\ (\ref{eq:ddf_general}) by adding a thermal noise term,
\begin{equation}
\label{eq:ddf_general2}
 \partial_t \phi_\ga = \nabla \int \ud \rr' \tLL_\ab(\rr,\rr') 
     \: \nabla' \tilde{\mu}_\gb(\rr') + \zeta_\ga(\rr,t),
\end{equation}
i.e., a stochastic Gaussian distributed field $\zeta_\ga(\rr,t)$ with 
zero mean ($\langle \zeta_\ga (\rr,t) \rangle = 0$),
which is correlated according to the fluctuation-dissipation theorem:
\begin{equation}
\label{eq:ddf_fluct}
\langle \zeta_\ga(\rr,t)\zeta_\gb(\rr',t') \rangle 
 = - 2 \frac{V}{n_\mr c} \: 
   \delta(t-t') \: \nabla \tLL_\ab(\rr,\rr') \nabla'
\end{equation}
Equivalently, one can add a fluctuating current,
\begin{equation}
\label{eq:ddf_general3}
 \partial_t \phi_\ga = \nabla \Big( \int \ud \rr' \tLL_\ab(\rr,\rr') 
     \: \nabla' \tilde{\mu}_\gb(\rr') + {\mb j}_\ga \Big),
\end{equation}
with $\langle {\mb j}(\rr,t) \rangle = 0$ and
\begin{equation}
\label{eq:ddf_fluct2}
\langle j_{\ga  I}(\rr,t) j_{\gb J}(\rr',t') \rangle 
 =  2 \: \frac{V}{n_\mr c} \: 
   \delta(t-t') \: \delta_{IJ} \: \tLL_\ab(\rr,\rr'), 
\end{equation}
where $I,J$ denote Cartesian coordinates. The amplitude of thermal noise 
is measured by the inverse Ginzburg parameter \cite{polymer_book}
$C^{-1} = k_B T \: V / n_\mr c R_g^3$. If $C^{-1}$ is large (e.g. $C^{-1}\gg 1$), fluctuations are 
important, and stochastic DDFT descriptions are required.

\end{appendix}


\begin{thebibliography}{99}

\section*{References}

\bibitem{morphology_book}
Guo, Q. 
\textit{Polymer Morphology: Principles, Charcterization, and Processing},
Wiley, 2016.

\bibitem{multiphase_book}
Boudenne, A.; Ibos, L., Candau, Y.; Thomas, S. Eds.,
\textit{Handbook of multiphase polymer systems},
Wiley, 2011.

\bibitem{scattering_spinodal_decomposition}
 Bailey, A. E.; Poon, W. C. K.; Christianson, R. J.; Schofield, A. B.; Gasser, U.; Prasad, V.; Manley, S.; Segre, P. N.; Cipelletti, L.; Meyer, W. V.; Doherty, M. P.; Sankaran, S.; Jankovsky, A. L.; Shiley, W. L.; Bowen, J. P.; Eggers, J. C.; Kurta, C.; Jr. Lorik, T.; Pusey, P. N.; Weitz, D. A. 
Spinodal decomposition in a model colloid-polymer mixture in microgravity.
\textit{Phys. Rev. Lett.} \tb{2007}, 99, 205701. 

\bibitem{neutron_interface_formation}
B\'eziel, W.; Fragneto, G.; Cousin, F.; Sferrazza, M. 
Neutron reflectivity study of the kinetics of polymer-polymer interface formation.
\textit{Phys. Rev. E} \tb{2008}, 78, 022801.

\bibitem{optoelectronics_review} 
Pearson, A. J.; Wang, T.; Lidzey, D. G.
The role of dynamic measurements in correlating structure with
optoelectronic properties in polymer:fullerene bulk-heterojunction
solar cells.
\textit{Rep. Progr. Phys.} \tb{2013}, 76, 022501.

\bibitem{micelle_fusion}Zhang, C.; Fan, Y.; Zhang, Y.; Yu, C.; Li, H.; Chen, Y.; Hamley, I. W.; Jiang, S. 
Self-assembly kinetics of amphiphilic dendritic copolymers. 
\textit{Macromolecules} {\bf 2017}, 50, 1657-1665.

\bibitem{Kawasaki1}
Kawasaki, K.; Sekimoto, K. Dynamical theory of polymer melt morphology. 
\textit{Physica A} \tb{1987}, 143, 349--413.

\bibitem{Kawasaki2}
Kawasaki, K.; Sekimoto, K. 
Morphology dynamics of block copolymer systems.  
\textit{Physica A } \tb{1988}, 148, 361--413.

\bibitem{Harden} 
Harden, J. 
Kinetics of interface formation between weakly incompatible polymer blends. 
\textit{J. Physique} \tb{1990}, 51, 1777--1784.

\bibitem{dynamic_MF}
Fraaije, J. G. E. M.; van Vlimmeren, B. A. C.; Maurits, N. M.; Postma, M.; Evers, O. A.; Hoffmann, C.; Altevogt, P.; Goldbeck-Wood, G. 
The dynamic mean-field density functional method and its application to the mesoscopic dynamics of quenched block copolymer melts. 
\textit{J. Chem. Phys.} {\bf 1997}, 106, 4260-4269.

\bibitem{EPD1997}
Maurits, N. M.; Fraaije, J. G. E. M. 
Mesoscopic dynamics of copolymer melts: from density dynamics to external potential dynamics using nonlocal kinetic coupling. 
\textit{J. Chem. Phys.} \tb{1997}, 107, 5879-5889

\bibitem{Doi_adsorption}
Hasegawa, R.; Doi, M. Adsorption dynamics. 
Extension of self-consistent field theory to dynamic problems. 
\textit{Macromolecules} {\bf 1997}, 30, 3086-3089.

\bibitem{Kawakatsu}
Kawakatsu, T.
Effects of changes in the chain conformations on the kinetics of
order-disorder transitions in block copolymer melts.
\textit{Phys. Rev. E} \tb{1997}, 56, 3240-3246.

\bibitem{Shi_interface}
Yeung, C.; Shi, A.-C. 
Formation of interfaces in incompatible polymer blends: a dynamical mean field study.
\textit{Macromolecules} {\bf 1999}, 32, 3637-3642.

\bibitem{Shima} 
Shima, T.; Kuni, H.; Okabe, Y.; Doi, M.; Yuan, X.-F.; Kawakatsu, T. 
Self-consistent field theory of viscoelastic behavior of inhomogeneous dense polymer systems.
\textit{Macromolecules} \tb{2003}, 36, 9199--9204.

\bibitem{fluctuation_dynamics}
M\"uller, M.; Schmid, F. 
Incorporating fluctuations and dynamics in self-consistent field theory for polymer blends. 
\textit{Adv. Polym. Sci.} \tb{2005}, 185, 1-58.

\bibitem{EPD_vesicle}
He, X.; Schmid, F. 
Dynamics of spontaneous vesicle formation in dilute solutions of amphiphilic diblock copolymers. 
\textit{Macromolecules} {\bf 2006}, \tb{39}, 2654-2662.

\bibitem{DFT_vesicle}
Uneyama, T. Density functional simulation of spontaneous formation of vesicle in block copolymer solutions. 
\textit{J. Chem. Phys.} \tb{2007}, 126, 114902.

\bibitem{spinodal_critical}
Zhang, X.; Qi, S.; Yan, D. 
Spinodal assisted growing dynamics of critical nucleus in polymer blends. 
\textit{J. Chem. Phys.} \tb{2012}, 137, 184903.

\bibitem{book_cmp}
Chaikin, P. M.; Lubensky, T. C. 
\textit{Principles of condensed matter physics}, 
Cambridge University Press: 1995.

\bibitem{dynamic_review}
Hohenberg, P. C.; Halperin, B. I. 
Theory of dynamic critical phenomena. 
\textit{Rev. Mod. Phys.} \tb{1977}, 49, 436-479.

\bibitem{ohta_kawasaki}
Ohta, T.; Kawasaki, K.
Equilibrium morphology of block copolymer melts.
\textit{Macromolecules} \tb{1986}, 19, 2621--2632.

\bibitem{gomez_15}
Gomez, L. R.; Garica, N. A.; Vitelli, V.; Lorenzana, J.; Vega, D.
Phase nucleation in curved space.
\textit{Nature Communications} \tb{2015}, 6, 6856.

\bibitem{abate_16}
Abate, A.; Vu, G.; Pezzutti, A.; Garcia, N.; Davis, R.; Schmid, F.; 
Register, R.; Vega, D.
Shear-aligned block copolymer monolayers as seeds to control the 
orientational order in cylinder forming block copolymer thin films.
\textit{Macromolecule} \tb{2016}, 49, 7588.

\bibitem{simon_16}
Kessler, S.; Schmid, F.; Drese, K.
Modeling size controlled nanoparticle precipitation with the co-solvency
method by spinodal decomposition.
\textit{Soft Matter} \tb{2016}, 12, 7231--7240.

\bibitem{helfand_75}
Helfand, E.
Theory of inhomogeneous polymers -- fundamentals of Gaussian random-walk model.
\textit{J. Chem. Phys.} \tb{1975}, 62, 999--1005.

\bibitem{freed_95}
Freed, K.F.
Interrelation between density-functional and self-consistent-field
formulations for inhomogeneous polymer systems.
\textit{J. Chem. Phys.} \tb{1995}, 103, 3230--3239.

\bibitem{review_scf}
Schmid, F.
Self-consistent field theories for complex fluids.
\textit{J. Phys.: Cond. Matter} \tb{1998}, 10, 8105--8138.

\bibitem{matsen_review}
Matsen, M.W.
The standard Gaussian model for block copolymer melts.
\textit{J. Phys.: Cond. Matter} \tb{2002}, 14, R21-R47.

\bibitem{review2_scf}
Schmid, F.
Theory and simulation of multiphase polymer systems.
in \textit{Handbook of multiphase polymer systems}
Boudenne, A.; Ibos, L., Candau, Y.; Thomas, S. Eds.,
chapter 3, 31--80, Wiley 2011.

\bibitem{Dean1996}
Dean, D. S. Langevin equation for the density of a system of interacting Langevin processes. 
\textit{J. Phys. A: Math. Gen.} {\bf 1996}, 29, L613-L617.

\bibitem{SDFT}
Frusawa, H; Hayakawa, R. 
On the controversy over the stochastic density functional equations. 
\textit{J. Phys. A: Math. Gen.} {\bf 2000}, 33, L155-L160.

\bibitem{DFT_fluid}
Marconi, U. M. B.; Tarazona, P. 
Dynamic density functional theory of fluids. 
\textit{J. Chem. Phys.} {\bf 1999}, 110, 8032-8044.

\bibitem{DFT_spinodal}
Evans; R, Archer, A. J. 
Dynamic density functional theory and its application to spinodal decomposition.
\textit{J. Chem. Phys.} {\bf 2004}, 121, 4246-4254.

\bibitem{DFT_freezing}
Bagchi, B. 
Stability of supercooled liquid to periodic density waves and dynamics of freezing. 
\textit{Physica A}, {\bf 1987}, 145, 273-289.


\bibitem{DDFT_SD}
Archer, A. J.; Rauscher, M. 
Dynamic density functional theory for interacting Brownian particles: stochastic or deterministic. 
\textit{J. Phys. A: Math. Gen.} \tb{2004}, 37, 9325.


\bibitem{power_functional2}
Brader, J. M.; Schmidt. M.
Power functional theory for the dynamic test particle limit.
\textit{J. Phys.: Cond. Matter} \tb{2015}, 27, 194106.

\bibitem{power_functional}
Schmidt, M.; Brader, J. M.
Power functional theory for Brownian dynamics.
\textit{J. Chem. Phys.} \tb{2013}, 138, 214101.

\bibitem{DFT_JPCM}
Marconi, U. M. B.; Tarazona, P. 
Dynamic density functional theory of fluids. 
\textit{J. Phys.: Condens. Matter} {\bf 2000}, 12, A413-A418.

\bibitem{dynamics_PI_SCMF}
Fredrickson, G. H.; Orland, H. Dynamics of polymers: a mean-field theory. 
\textit{J. Chem. Phys.} {\bf 2014}, 140, 084902

\bibitem{dynamics_path_integral}
Grzetic, D. J.; Wickham, R. A.; Shi, A.-C. 
Statistical dynamics of classical systems: a self-consistent field approach. 
\textit{J. Chem. Phys.} {\bf 2014}, 140, 244907.

\bibitem{ganesan1}
Ganesan, V.; Pryamitsyn, V. A.
Dynamical mean-field theory for inhomogeneous polymeric systems.
\textit{J. Chem. Phys.} \tb{2003}, 118, 4345-4348.

\bibitem{interface_SCFBD}
Narayanan, B.; Pryamitsyn, V. A.; Ganesan, V. 
Interfacial phenomena in polymer blends: a self-consistent Brownian dynamics study. 
\textit{Macromolecules} \tb{2004}, 37, 10180-10194.

\bibitem{scmf1}
M\"uller, M.; Smith, G. D.
Phase separation in binary mixtures containing polymers: A quantitative
comparison of single-chain-in-mean-field simualtions and computer simulations
of the corresponding multichain system.
\textit{J. Polym. Sci., Part B} \tb{2005}, 43, 934--958.

\bibitem{scmf2}
Daoulas, K. Ch.; M\"uller, M.
Single-chain in mean field simulations: Quasi-instantaneous field approximation
and quantitative comparison with Monte Carlo simulations.
\textit{J. Chem. Phys.} \tb{2006}, 125, 184904.

\bibitem{hybrid_MD}
Milano, G.; Kawakatsu, T. 
Hybrid particle-field molecular dynamics simulations for dense polymer systems. 
\textit{J. Chem. Phys.} \tb{2009}, 130, 214106.

\bibitem{Zwanzig}
Zwanzig, R. 
\textit{Nonequilibrium statistical mechanics}, 
Oxford University Press, 2001.

\bibitem{projection1}
Kinjo, T.; Hyodo, S.-A. 
Equation of motion for coarse-grained simulation base on microscopic description. 
\textit{Phys. Rev. E} {\bf 2007}, 75, 051109.

\bibitem{projection2}
Hij\'on, C.; Espa\~nol, P.; Vanden-Eijnden, E.; Delgado-Buscalioni, R. 
Mori-Zwanzig formalism as a practical computational tool. 
\textit{Faraday Discuss.} {\bf 2010}, 144, 301-322. 

\bibitem{EPD_spinodal}
Reister, E.; M\"uller, M.; Binder, K. 
Spinodal decomposition in a binary polymer mixture: dynamic self-consistent-field theory and Monte Carlo simulations. 
\textit{Phys. Rev. E} \tb{2001}, 64, 041804.

\bibitem{EPD_layer}
Reister, E.; M\"uller, M. 
Formation of enrichment layers in thin polymer films: the influence of single chain dynamics. 
\textit{J. Chem. Phys.} \tb{2003}, 118, 8476-8488

\bibitem{EPD_micelle}
He, X. H.; Schmid, F.
Spontaneous formation of complex micelles from homogeneous solution.
\textit{Phys. Rev. Lett.} \tb{2008}, 100, 137802.

\bibitem{interface_poly}
Qi, S.; Zhang, X.; Yan, D. External potential dynamic studies on the formation of interface in polydisperse polymer blends. 
\textit{J. Chem. Phys.} \tb{2010}, 132, 064903

\bibitem{EPD_fields}
Fan, J.J.; Li, W.; Pan, D.; Shi, M. F.
External potential dynamics simulations of morphological transitions in diblock copolymer melt under an electric field.
\textit{Adv. Mater. Res.} \tb{2014}, 915-916, 545--548.

\bibitem{EPD_composites}
Raman, V.; Hatton, T. A.; Olsen, B. D.
Kinetics of magnetic field-induced orientational ordering in block copolymer /
superparamagnetic nanoparticle composites.
\textit{Macromolecular Rapid Communications} \tb{2014}, 35, 2005--2011.

\bibitem{EPD_T}
Li, W.; Jiang, W.
External potential dynamics simulation of the compatibility of T-shaped
graft copolymer compatibilizing two immiscible homopolymers.
\textit{E-polymers} \tb{2010}, 055

\bibitem{heuser}
Heuser, J. ; Sevink, G.J.A.; Schmid, F.
Self-assembly of polymeric particles in Poiseuille flow: A hybrid
Lattice Boltzmann/External Potential Dynamics simulation study.
\textit{Macromolecules} \tb{2017}, 
doi:10.1021/acs.macromol.6b02684.

\bibitem{brush_switch}
Qi, S.; Klushin, L. I.; Skvortsov, A. M.; Polotsky, A. A.; Schmid, F. 
Stimuli-responsive brushes with active minority components: Monte Carlo study an analytical theory. 
\textit{Macromolecules} \tb{2015}, 48, 3775-3787.

\bibitem{Edwards}
Edwards, S. F. 
The statistical mechanics of polymers with excluded volume. 
\textit{Proc. Phys. Soc.} {\bf 1965}, 85, 613-624.

\bibitem{brush1994}
Laradji, M.; Guo, H.; Zuckerman, M. J. 
Off-lattice Monte Carlo simulation of polymer brushes in good solvents. 
\textit{Phys. Rev. E} \tb{1994}, 49, 3199-3209.

\bibitem{hybrid_solution}
Qi, S.; Behringer, H.; Raasch, T.; Schmid, F. 
A hybrid particle-continuum resolution method and its application to a homogeneous solution. 
\textit{Eur. Phys. J. Special Topics} \tb{2016}, 225, 1527-1549.

\bibitem{CIC}
Birdsall, C. K.; Fuss, D.
Clouds-in-Clouds, Clouds-in-Cell Physics for Many-Body Plasma Simulations.
\textit{J. Comput. Phys.} \tb{1997}, 135, 141--148.

\bibitem{stochastic_numerical}
Leimkuhler, B.; Matthews, C. 
Robust and efficient configurational molecular sampling via Langevin dynamics. 
\textit{J. Chem. Phys.} \tb{2013}, 138, 174102.

\bibitem{EdwardsType2}
Besold, G.; Guo, H.; Zuckermann, M. J. 
Off-lattice Monte Carlo simulation of discrete Edwards model. 
\textit{J. Polym. Sci., Part B: Polym. Phys.} \tb{2000}, 38, 1053-1068

\bibitem{particle_mesh}
Detcheverry, F. A.; Kang, H.; Daoulas, K. Ch.; M\"uller, M.; Nealey, P. F.; de Pablo, J. J. 
Monte Carlo simulations of a coarse grain model for block copolymers and nanocomposites. 
\textit{Macromolecules} \tb{2008}, 41, 4989-5001.

\bibitem{hybrid}
Qi, S.; Behringer, H.; Schmid, F. 
Using field theory to construct hybrid particle-continuum simulation schemes with adaptive resolution for soft matter systems. 
\textit{New. J. Phys.} \tb{2013}, 15, 125009.

\bibitem{Sevink}
Sevink, G.J.A.; Schmid, F.; Kawakatsu, T.; Milano, G.
Combining cell-based hydrodynamics with hybrid particle-field simulations:
Efficient and realistic simulation of structuring dynamics.
\textit{Soft Matter} \tb{2017}, 13, 1594-1623.

\bibitem{Doi_book}
Doi, M.; Edwards, S. F.
\textit{The theory of polymer dynamics},
Clarendon Press, 1988.

\bibitem{polymer_book}Fredrickson, G. H. 
\textit{The equilibrium theory of inhomogeneous polymers}, 
Oxford University Press: 2006.

\bibitem{semi_implicit}Ceniceros, H. D.; Fredrickson, G. H. Numerical solution of polymer self-consistent field theory. \textit{Multiscale Model. Simul.} \tb{2004}, 2, 452-474.


\bibitem{Werner}
Werner, A.; Schmid, F.; M\"uller, M.; Binder, K.
``Intrinsic" profiles and capillary waves at homopolymer interfaces: A Monte Carlo study.
\textit{Phys. Rev. E} \tb{1999}, 59, 728-738.

\bibitem{interface_FH}
Wang, S.-Q; Shi, Q. 
Interdiffusion in binary polymer mixtures. 
\textit{Macromolecules} \tb{1993}, 26, 1091-1096

\bibitem{blend_E1}Chaturvedi, U. K.; Steiner, U.; Zak, O.; Krausch, G.; Klein, J. Interfacial structure in polymer mixtures below the critical point. \textit{Phys. Rev. Lett.} \tb{1989}, 63, 616-619.
\bibitem{blend_E2}Steiner, U.; Krausch, G.; Schatz, G.; Klein, J. Dynamics of mixing between partially miscible polymers. \textit{Phys. Rev. Lett.} \tb{1990}, 64, 1119-1121.

\bibitem{lamellar_ordering_1}Floudas, G.; Pakula, T.; Fischer, E. W.; Hadjichristidis, N.; Pispas, S. Ordering kinetics in a symmetric diblock copolymer. \textit{Acta Polymer.} \tb{1994}, 45, 176-181. 
\bibitem{lamellar_ordering_2}Floudas, G.; Vlassopoulos, D.; Pitsikalis, M.; Hadjichristidis, N.; Stamm, M. Order-disorder transition and ordering kinetics in binary diblock copolymer mixtures of styrene and isoprene. \textit{J. Chem. Phys.} \tb{1996}, 104, 2083-2088.
\bibitem{lamellar_ordering_3}Sakamoto, N.; Hashimoto, T. Ordering dynamics of a symmetric Polystyrene-block-polyisoprene. 2. Real-Space Analysis on the Formation of Lamellar Microdomain. \textit{Macromolecules} \tb{1998}, 31, 3815-3823.

\bibitem{selfassembly}
Grzybowski, B.A.; Wilmer, C.E.: Kim, J.: Browne, K.P.; Bishop, K.J.M.
Self-assembly: From crystals to cells.
\textit{Soft Matter} \tb{2009}, 5, 1110-1128.

\bibitem{Simon}Ke\ss ler, S; Drese, K; Schmid, F. Simulating copolymeric nanoparticle assembly in the co-solvent method: How mixing rates control final particle sizes and morphologies. \textit{Polymer} \tb{2017}, 126, 9-18.

\bibitem{Mueller_Tang}
M\"uller, M.; Tang, J.
Alignment of copolymer morphology by planar step elongation during spinodal self-assembly.
\textit{Phys. Rev. Lett} \tb{2015}, 115, 228301.

\end{thebibliography}
\end{document}